\documentclass[fleqn,usenatbib]{mnras}

\usepackage{makecell}
\usepackage{graphicx}    
\usepackage{amsmath}    
\usepackage{cases}
\usepackage{amssymb}    
\usepackage{multicol}   
\usepackage[utf8]{inputenc}
\usepackage{fancyref}
\usepackage{fullpage}
\usepackage{graphicx}
\usepackage{graphicx,natbib,url,twoopt}
\usepackage{float}
\usepackage{subfloat}
\usepackage{subfig}
\usepackage{caption}
\usepackage{xspace}
\usepackage{indentfirst}

\usepackage{lipsum}						
\usepackage{amsmath,amssymb,mathrsfs}	
\usepackage{setspace}  					
\usepackage{verbatim}					
\usepackage{tabularx} 					
\usepackage{afterpage} 					

\usepackage{romannum} 

\usepackage{graphicx}	
\usepackage{amsmath}	

\title[Short title, max. 45 characters]{Multifractality Signatures in Lensed Quasars}

\author[R. A. Assis Souza et al.]{
 R. A. Assis Souza$^{1,2}$, A. Bewketu Belete$^{3}$, B. L. Canto Martins$^{1}$, L. M. C. de Azevedo$^{1}$, \newauthor J. P. S. Campelo$^{1}$,  I. C Le\~ao$^{1}$, and J. R. De Medeiros$^{1}$ \\
$^{1}$Departamento de F\'isica Te\'orica e Experimental, Universidade Federal do Rio Grande do Norte, Natal RN, Brazil \\$^{2}$European Southern Observatory, Karl-Schwarzschild-Str. 2, 85748 Garching bei Munchen, Germany \\$^{3}$Department of Space, Earth and Environment (SEE), Chalmers Institute of Technology, Gothenburg, Sweden}

\date{Accepted October 2024. Received September 2023}

\pubyear{2024}

\begin{document}
\label{firstpage}
\pagerange{\pageref{firstpage}--\pageref{lastpage}}
\maketitle

\begin{abstract}
Variations in scaling behavior in the flux and emissions of gravitational lensed quasars can provide valuable information about the dynamics within the sources and their cosmological evolution with time. Here, we study the multifractal behavior of the light curves of 14 lensed quasars with multiple images in the $r$ band, with redshift ranging from 0.657 to 2.730, in the search for potential differences in nonlinearity between the signals of the quasar multiple images. Among these lensed systems, nine present two images, two present three images, and three present four images. To this end, we apply the wavelet transform-based multifractal analysis formalism called Wavelet Transform Modulus Maxima (WTMM). We identify strong multifractal signatures in the light curves of the images of all analyzed lensed quasar systems, independently of the number of images, with a significant difference between the degree of multifractality of all the images and combinations. We have also searched for a possible connection between the degree of multifractality and the characteristic parameters related to the quasar source and the lensing galaxy. These parameters include the Einstein ring radius and the accretion disk size and the characteristic timescales related to microlensing variability. The analysis reveals some apparent trends, pointing to a decrease in the degree of multifractality with the increase of the quasar's source size and timescale. Using a larger sample and following a similar approach, the present study confirms a previous finding for the quasar Q0957+561.
\end{abstract}

\begin{keywords}
methods: statistical  -- galaxies: active -- (galaxies:) quasars: general -- (galaxies:) quasars: emission lines
\end{keywords}



\section{Introduction} 
\label{sec:intro}

Lensed quasars are key laboratories for the study of several questions in Cosmology. Due to their high luminosity and variability, these objects are visible over cosmological distances and can be particularly useful for the study of time delay cosmography (\citealt{bonvin2016cosmograil,tewes2013cosmograil,millon2020cosmograil}). In addition, the referred lensing systems can be used in the study of microlensing effects and their properties (\citealt{mosquera2011microlensing,sluse2012microlensing,sun1989fitting}), the presence of cold dark matter (\citealt{richardson2022non},), and to study the properties of quasars themselves (\citealt{webster1992quasar}). Gravitational lensing effects can also be used to study the inner structure of lensed quasars (e.g., \citealt{jimenez2014average,braibant2017constraining,popovic2021spectroscopy}). Different emitting regions of lensed quasars can be affected differently by gravitational microlensing (e.g., \citealt{jovanovic2008microlensing} ), an aspect that can be used to investigate the accretion disc structure (e.g., \citealt{cornachione2020quasar}), as well as the structure and kinematics of the broad line region (BLR; see, e.g., \citealt{popovic2001influence,abajas2002influence,sluse2012microlensing,guerras2013microlensing,braibant2017constraining,hutsemekers2017new}). The analysis of quasars with different physical parameters, including a variety of redshifts, is also important for the study of the local environment effects on their evolution \citet{ellingson1991quasars}. For instance, different studies confirm the presence of quasars in relatively dense regions of galaxy clusters (\citealt{fisher1996galaxy,mclure2001cluster,barr2003cluster}).  

The study of the variability of quasars light curves (LCs) is based on different approaches for time-frequency analysis, which is associated with the physics to be unveiled behind the characteristics of the LCs. Among these approaches, the literature reports the autocorrelation function (ACF), the detrended fluctuation analysis (DFA), the multifractal detrended fluctuation analysis (MFDFA), the rescaled range statistical (R/S) analysis that provide type of self-affinity for stationary time series (\citealt{bashan2008comparison}), the periodogram regression (GPH) method, the (m, k)-Zipf method, and the detrended moving average (DMA) analysis ( \citealt{carbone2004analysis,shao2012comparing}). 

A series of papers \citet{belete2018multifractality,belete2019revealing,belete2019cosmological,belete2019novel,bewketu2020nature,belete2021molecular}  have searched for multifractal signatures in the light curves (LCs) of different quasars. For the lensed-quasar Q0957+561, in particular, \citet{belete2019revealing} have detected strong multifractal signature in the LCs of the images of the quasar Q0957+561, which changes over time monotonically, pointing to the presence of extrinsic variabilities in the LCs of the images. In an analysis of the LCs of the quasar Q0957+561, \citet{belete2019revealing} applied a wavelet transform-based multifractality approach called wavelet transform modulus maxima, which was first introduced by Muzy et al. (1991).

The present work aims to study the multifractal behavior in lensed quasars at different redshifts, following the same approach presented in \citet{belete2019revealing}. Using a large sample of lensed quasars with redshift ranging from 0.657 to 2.730, we search for potential differences in nonlinearity between the signals of the quasar multiple images and unravel their intrinsic variability and the extrinsic lensing mechanisms. For this purpose, we analyzed the multifractal (nonlinear) behavior of the LCs of multiple images of 14 lensed quasars in the $r$ band, using a wavelet transform-based multifractality analysis approach called Wavelet Transform Modulus Maxima (hereafter WTMM). We aim to identify potential signatures of multifractality, which reflects the difference in nonlinearity between the signals of the quasars' multiple images, and to verify whether there is a correlation between the degree of nonlinearity and the quasar's observational parameters. For this purpose, we analyzed the multifractal (nonlinear) behavior of the LCs of the multiple images of the quasars in the $r$ band using the WTMM. 

This paper is structured as follows. In Section \ref{sec:observations}, we discuss the data, method, and procedures used, and in Section \ref{sec:Results}, we present our results and discuss the multifractal nature of the 14 lensed quasars. We provide the summary and conclusions in Section \ref{sec:conclusions}.

\section{Working sample and data analysis} 
\label{sec:observations}

\subsection{Data collection}
\label{sec:datacollection}

\begin{figure*}
\begin{multicols}{3}
    \includegraphics[clip,width=\linewidth]{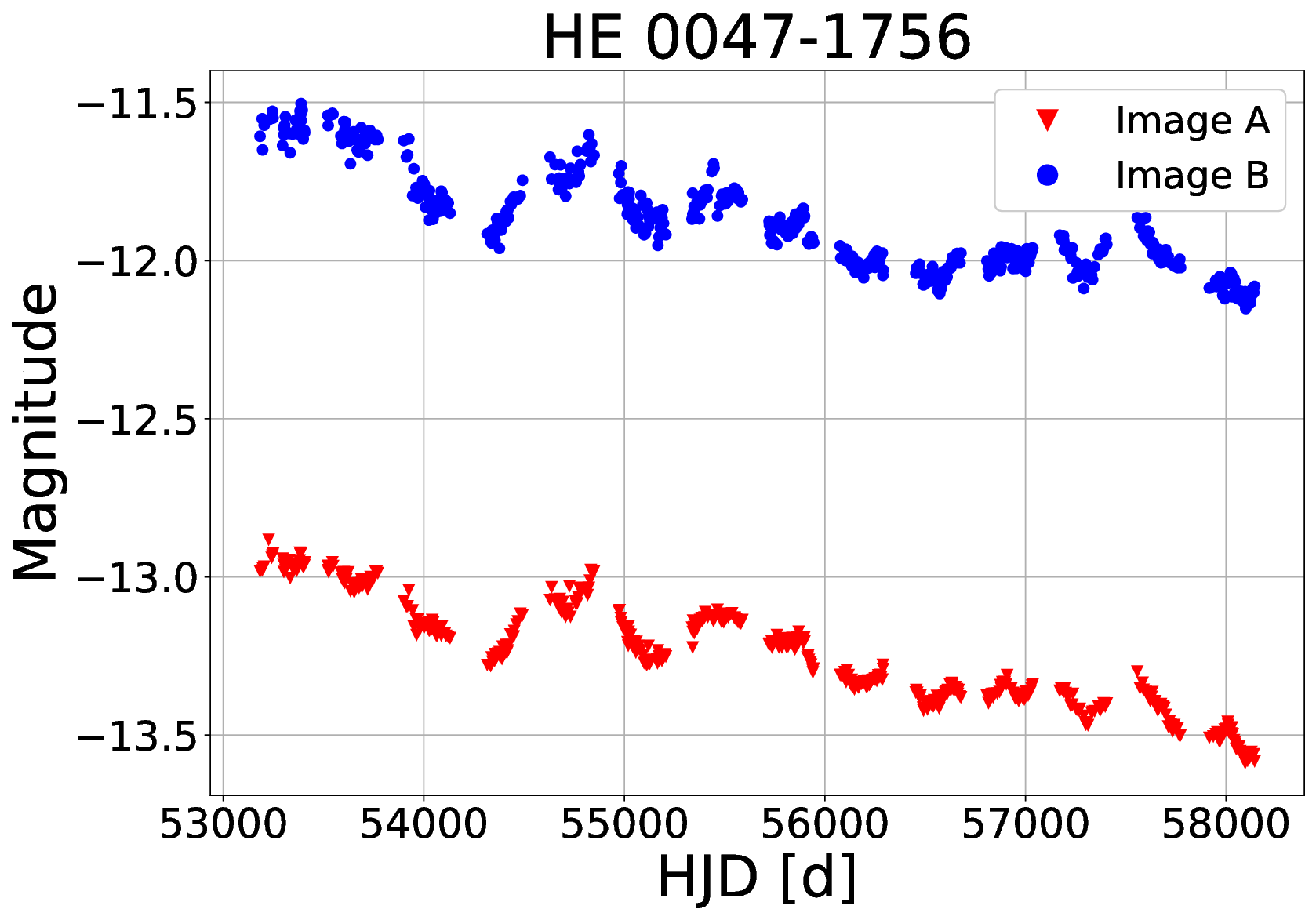}\par 
    \includegraphics[clip,width=\linewidth]{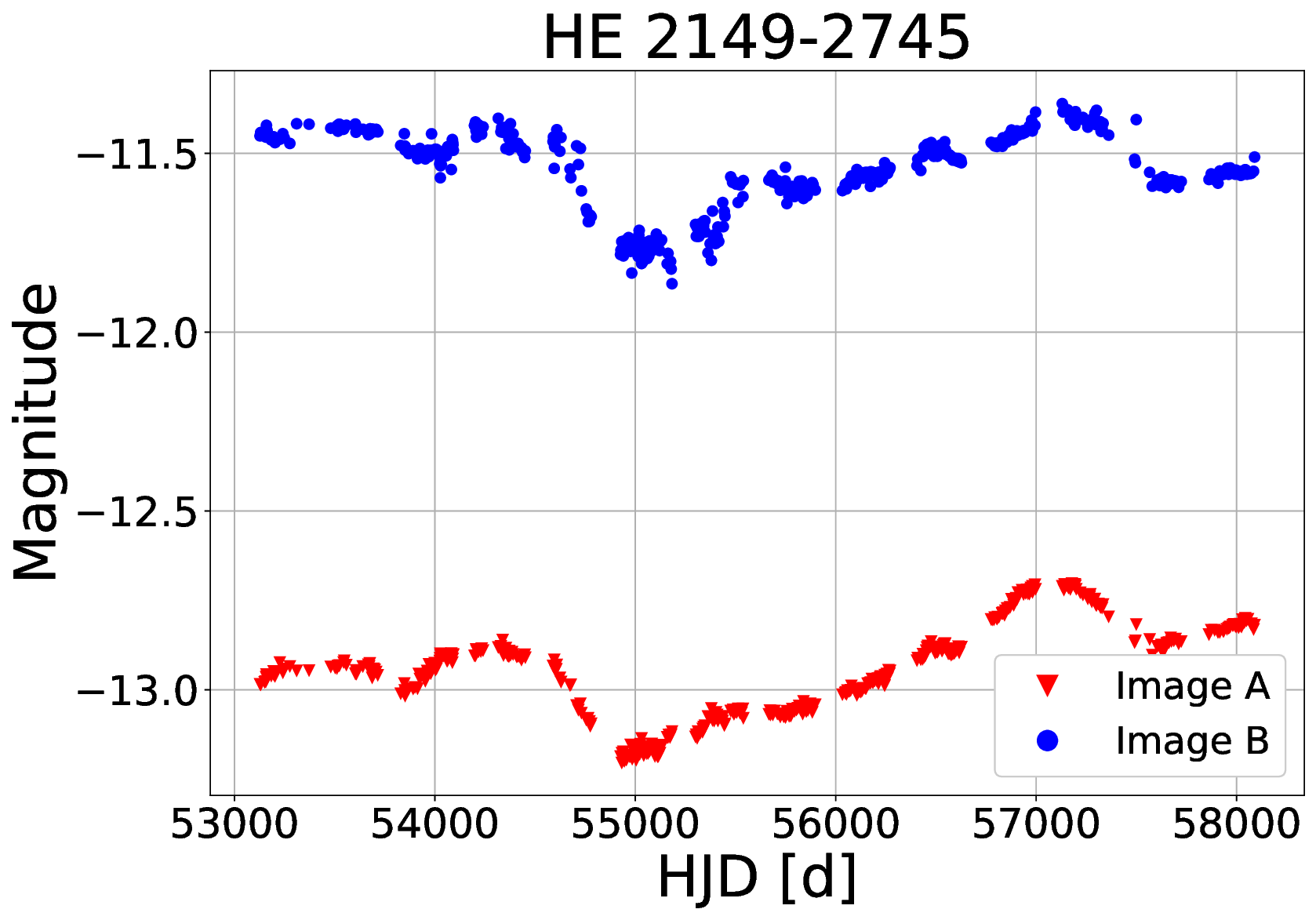}\par 
    \includegraphics[clip,width=\linewidth]{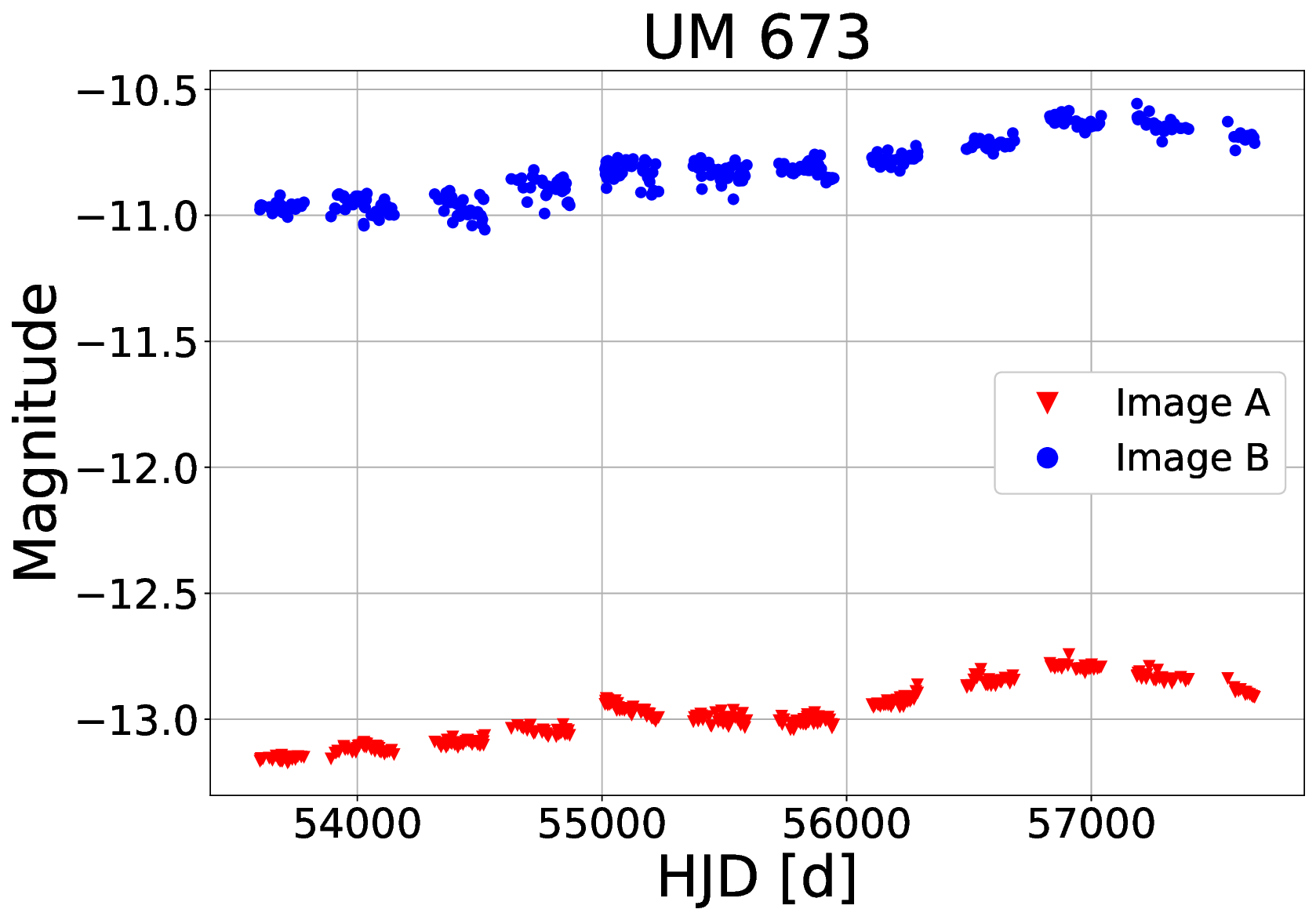}\par
    \includegraphics[clip,width=\linewidth]{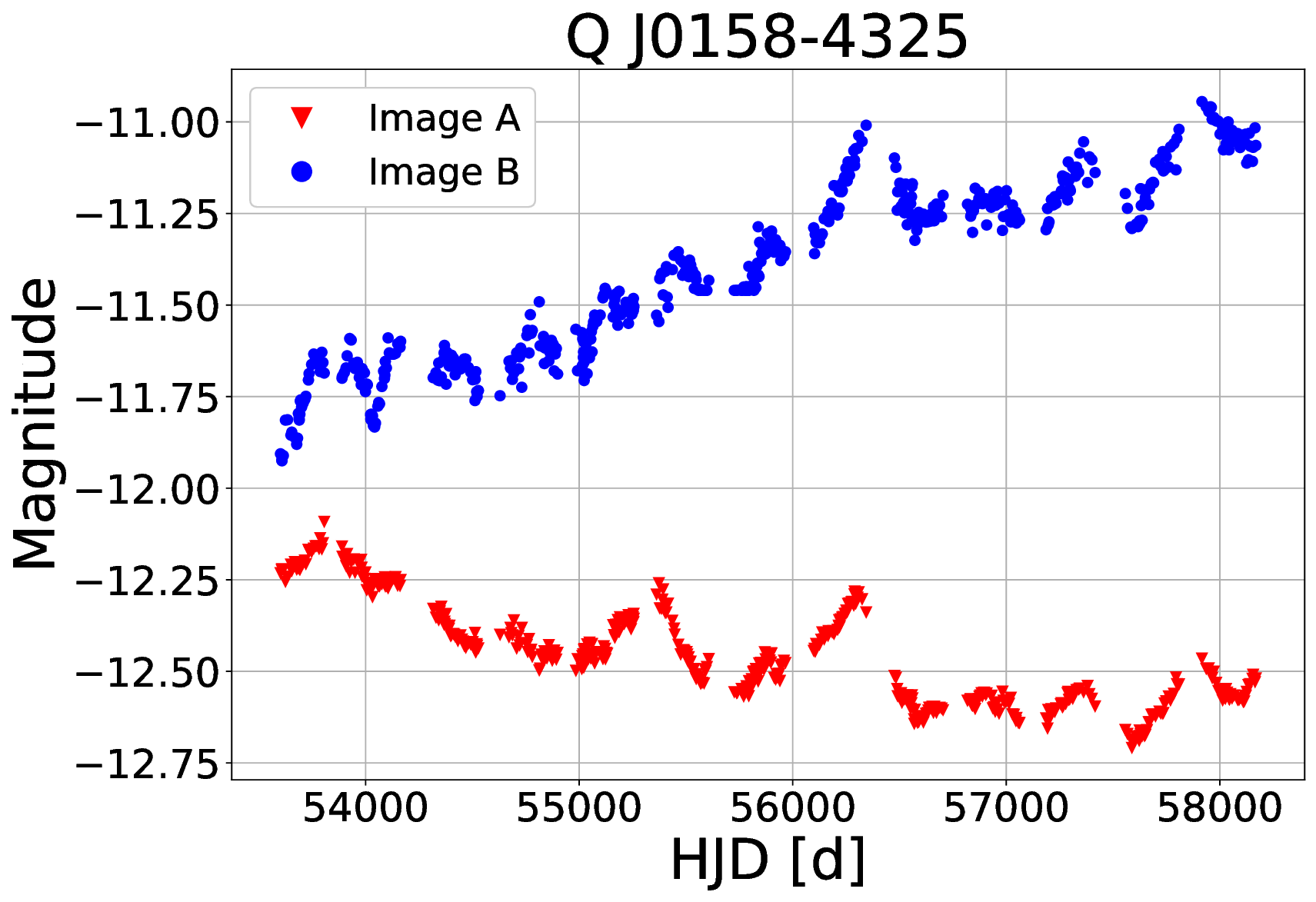}\par 
    \includegraphics[clip,width=\linewidth]{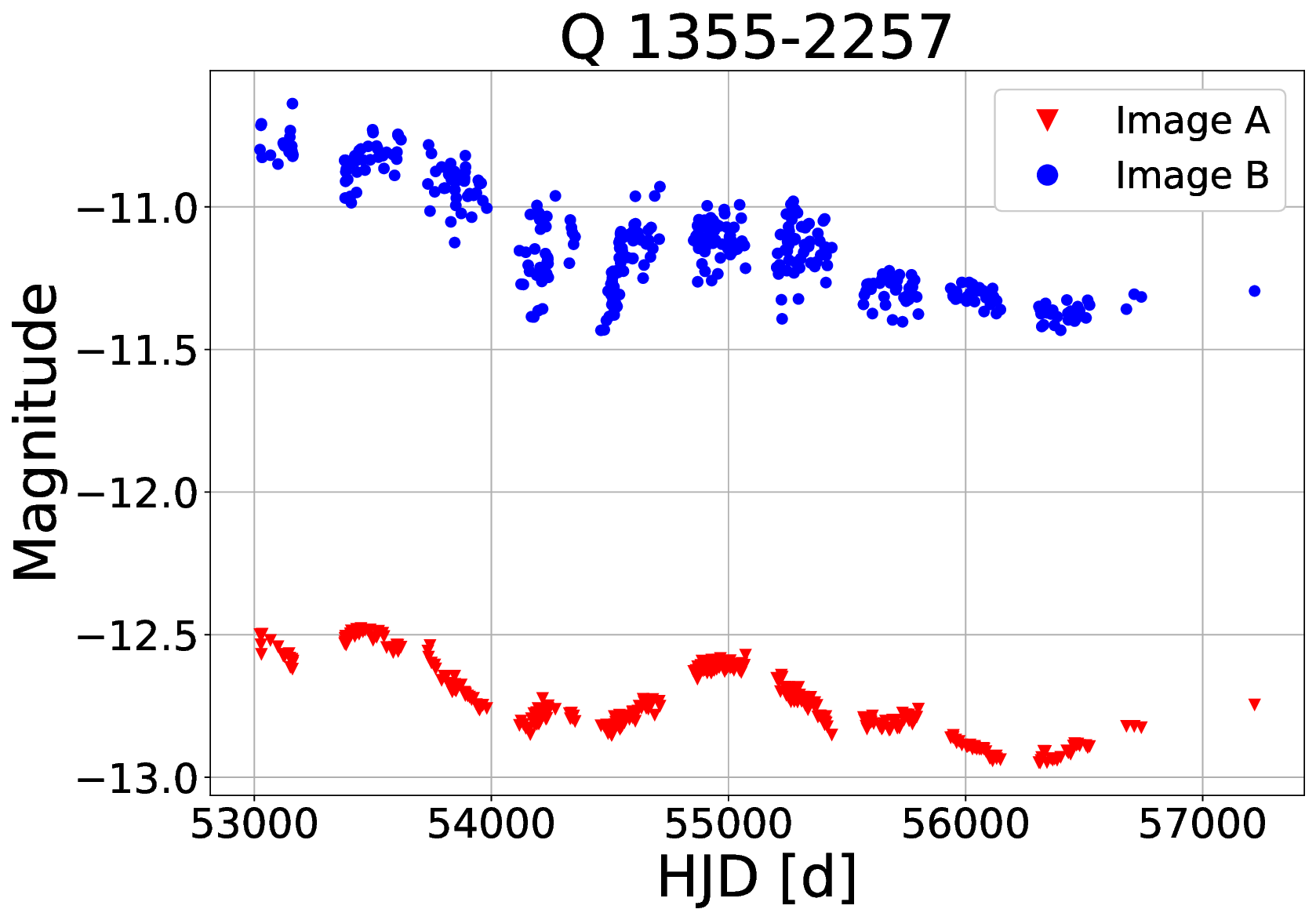}\par
    \includegraphics[clip,width=\linewidth]{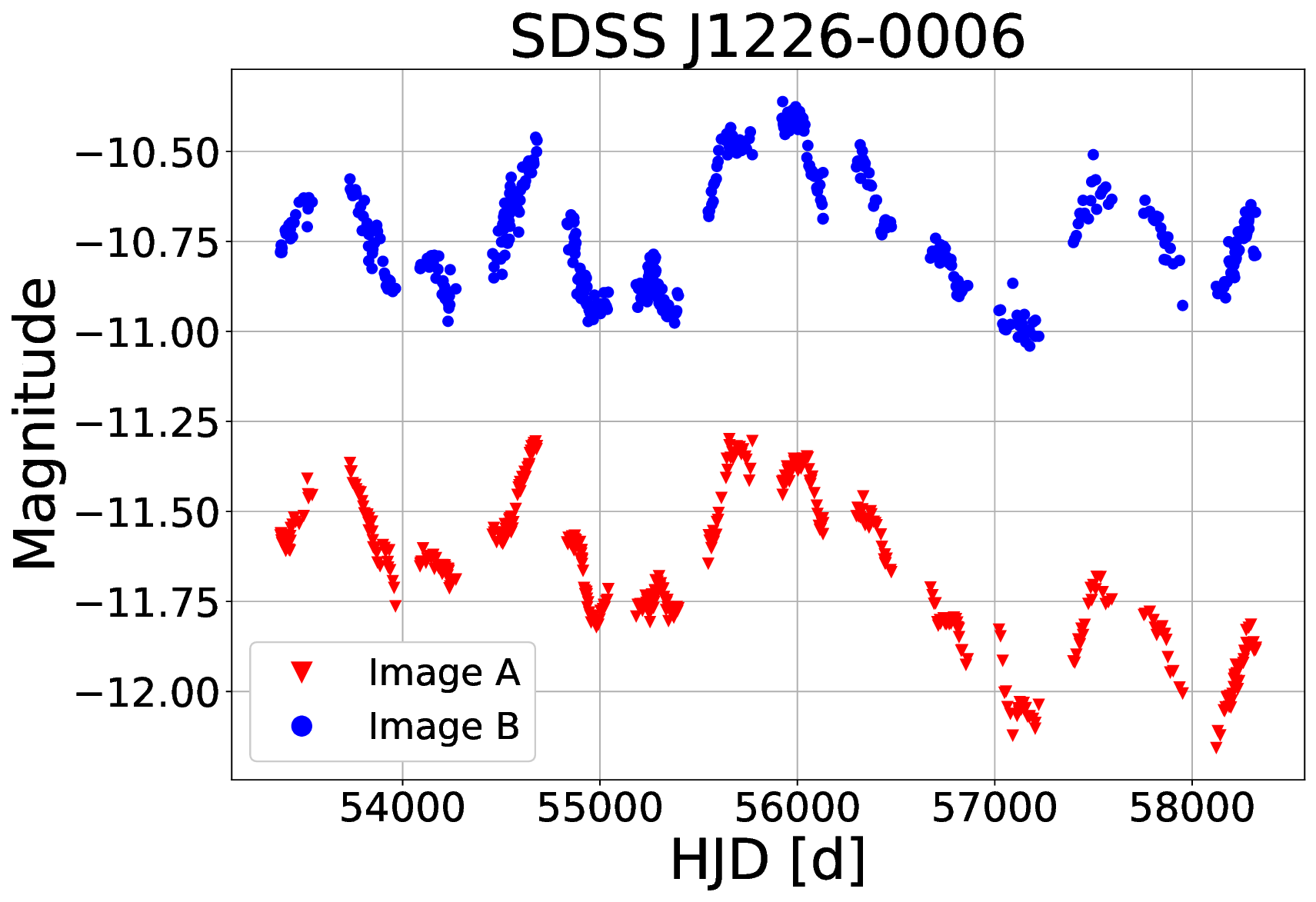}\par
    \includegraphics[clip,width=\linewidth]{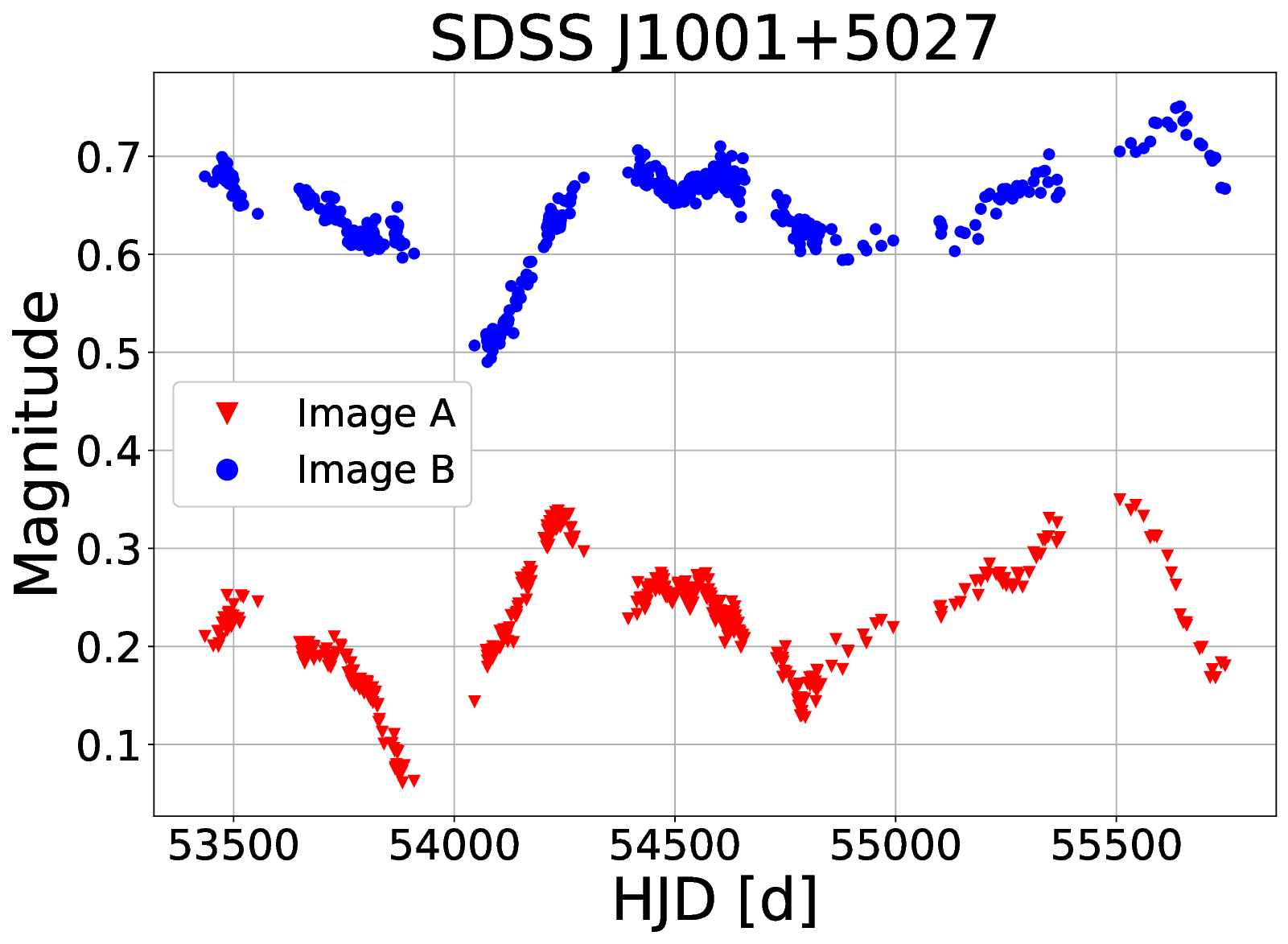}\par
    \includegraphics[clip,width=\linewidth]{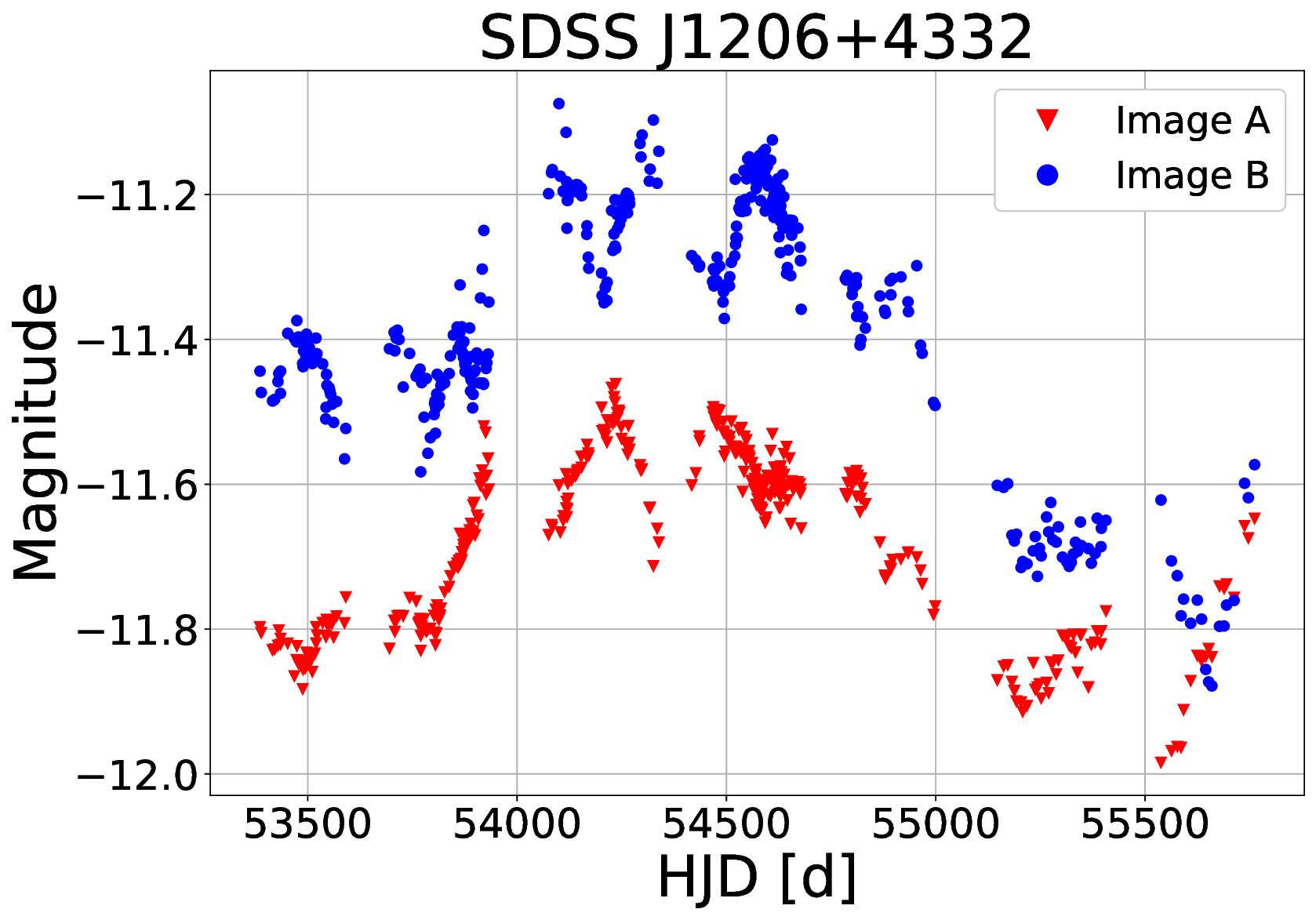}\par 
    \includegraphics[clip,width=\linewidth]{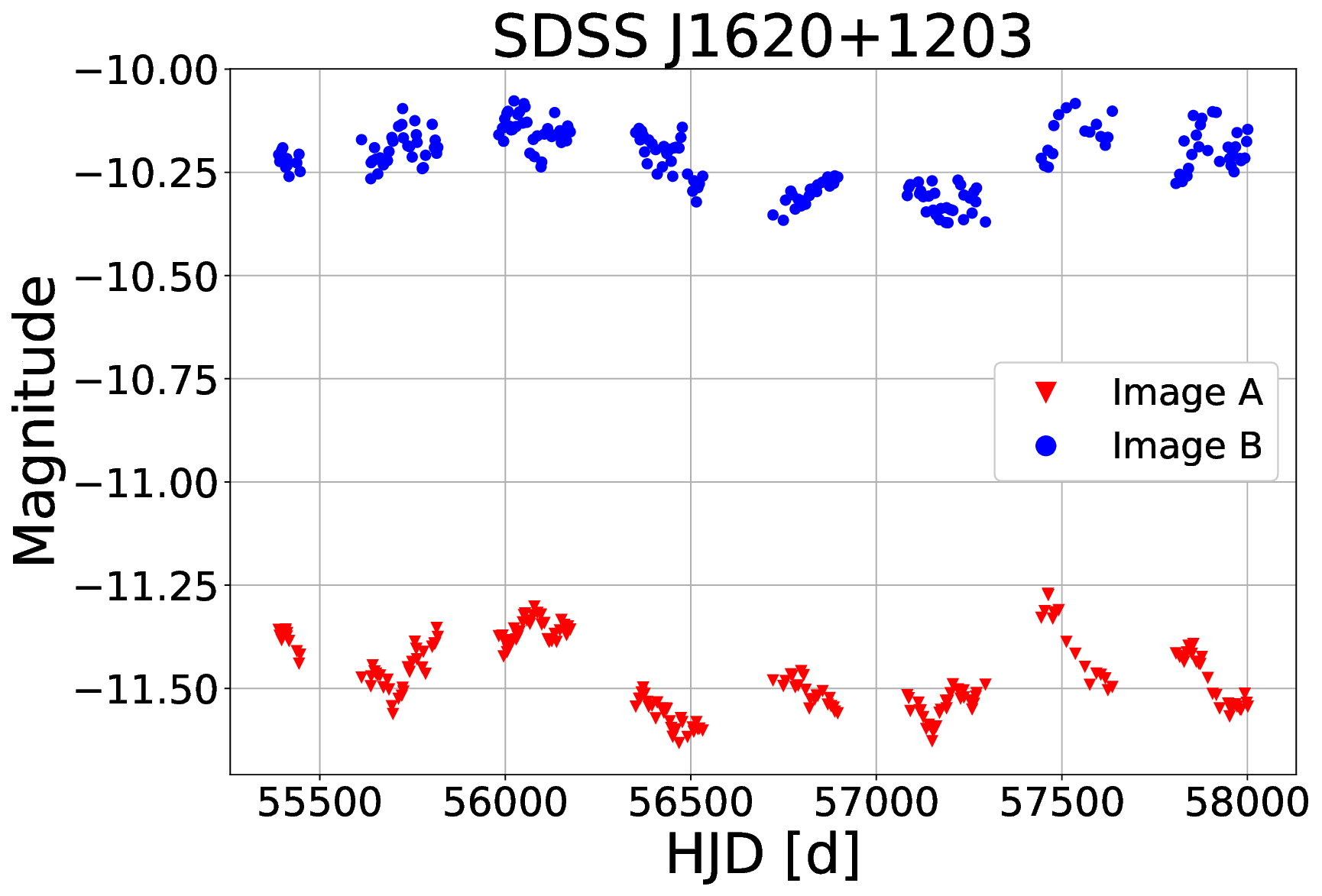}\par
    \includegraphics[clip,width=\linewidth]{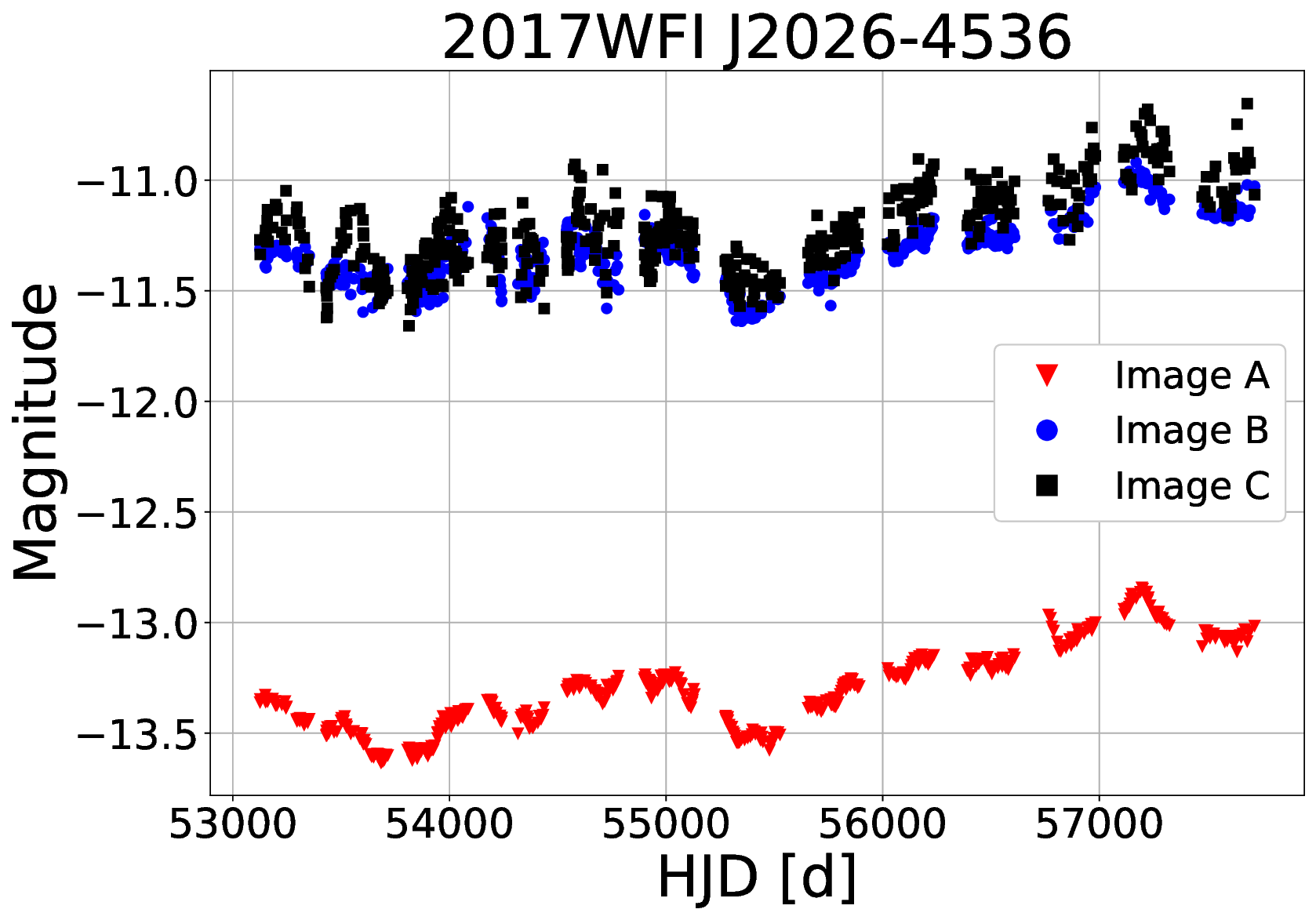}\par 
    \includegraphics[clip,width=\linewidth]{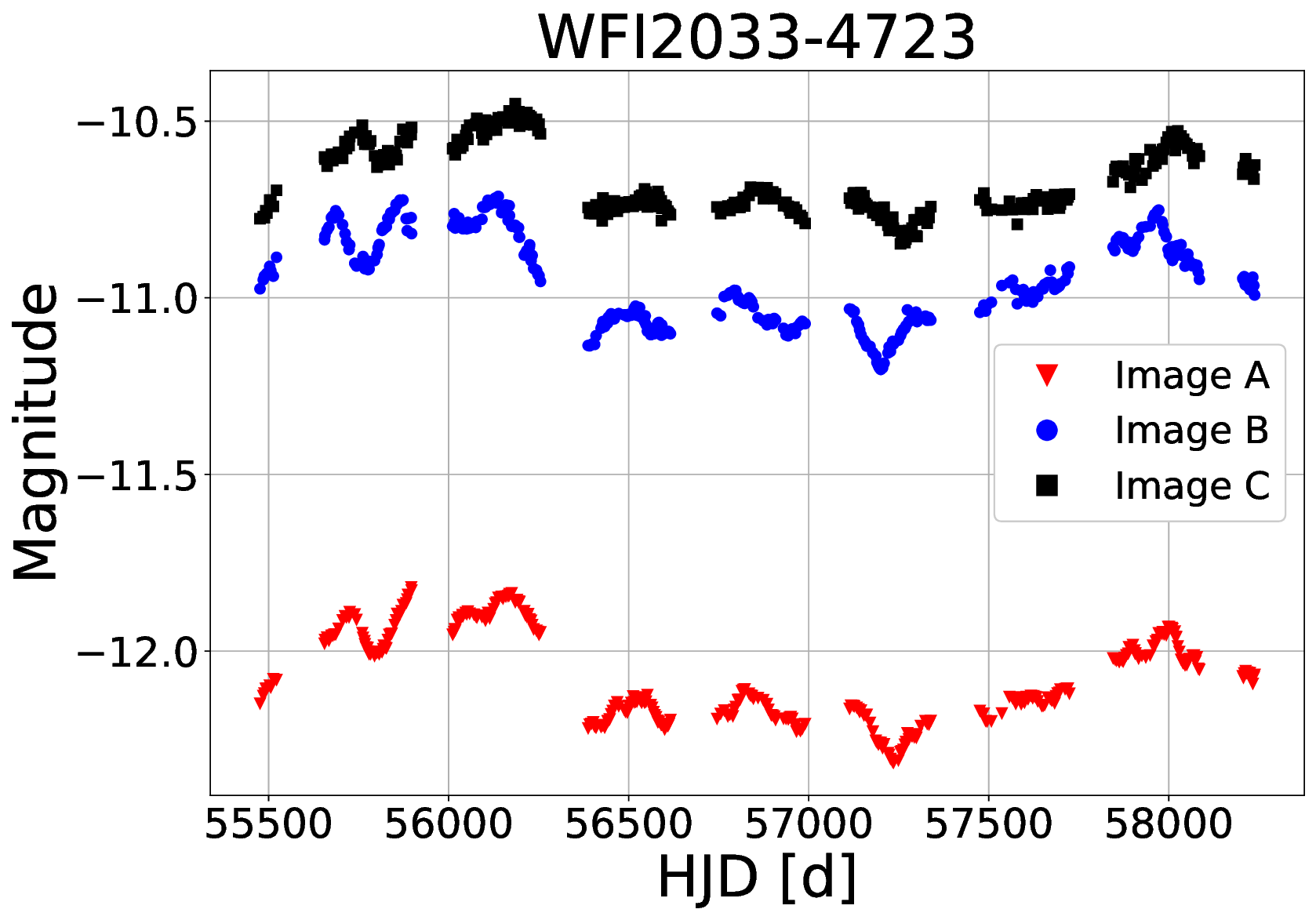}\par
    \includegraphics[clip,width=\linewidth]{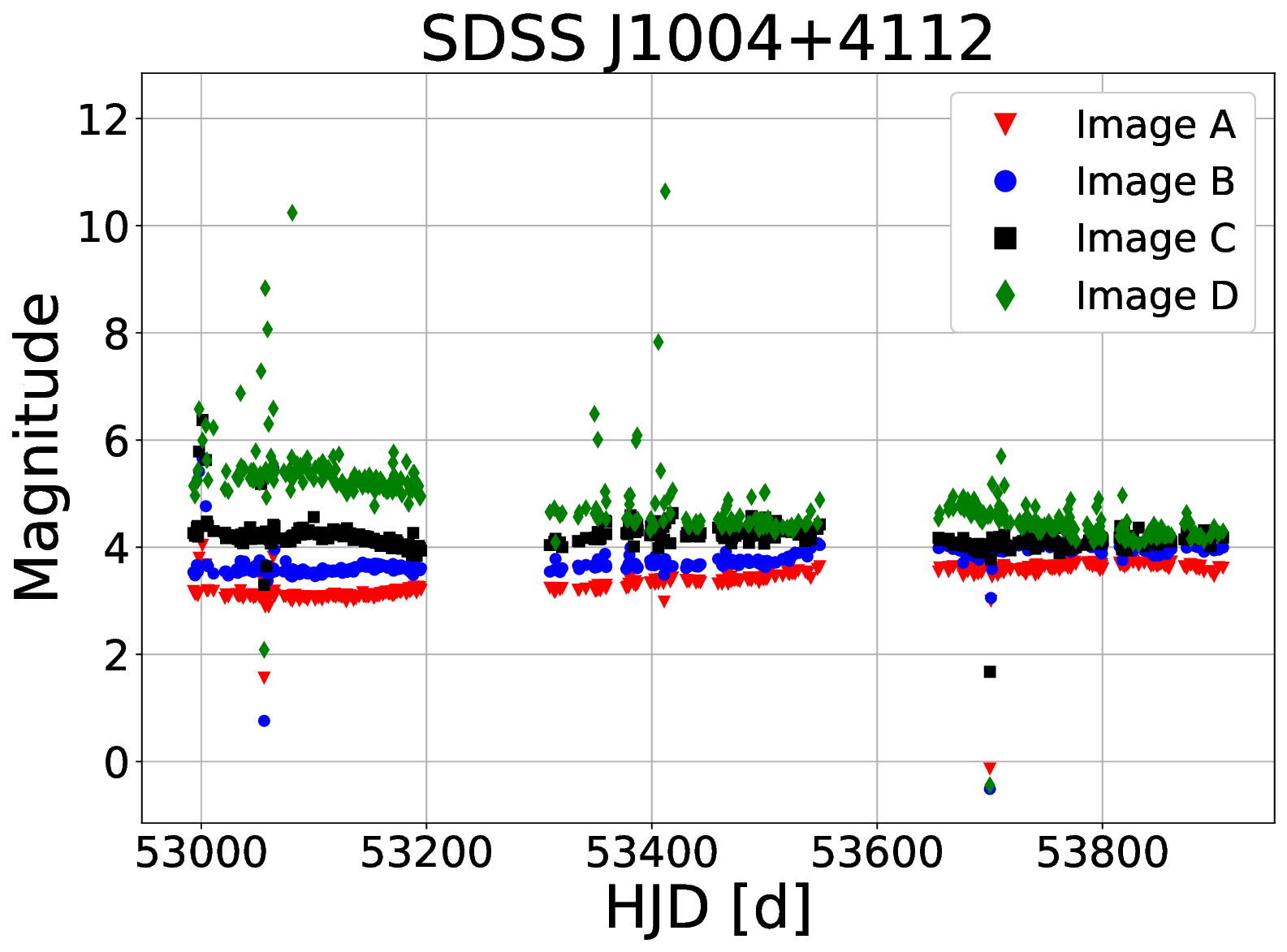}\par
    \includegraphics[clip,width=\linewidth]{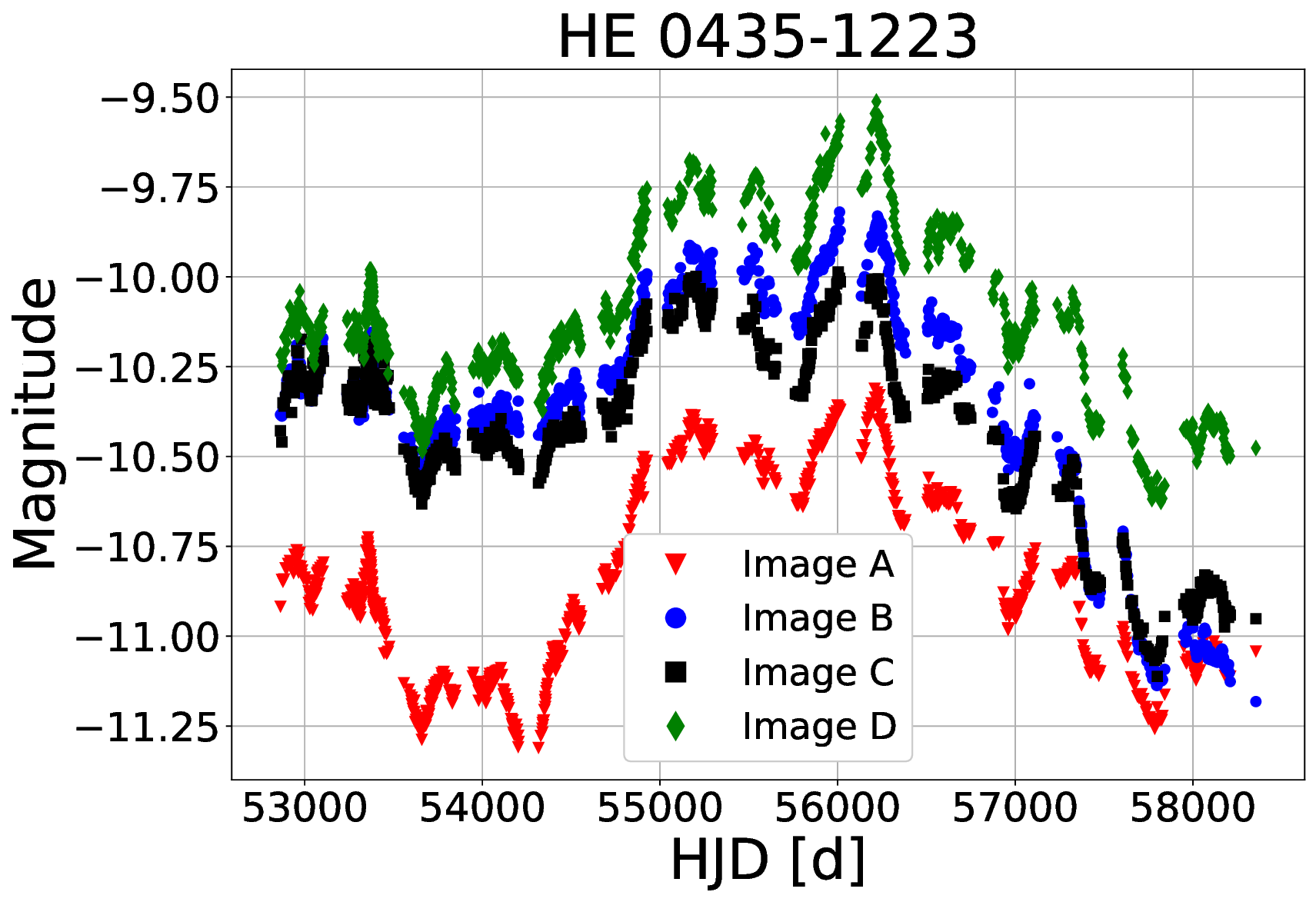}\par
    \includegraphics[clip,width=\linewidth]{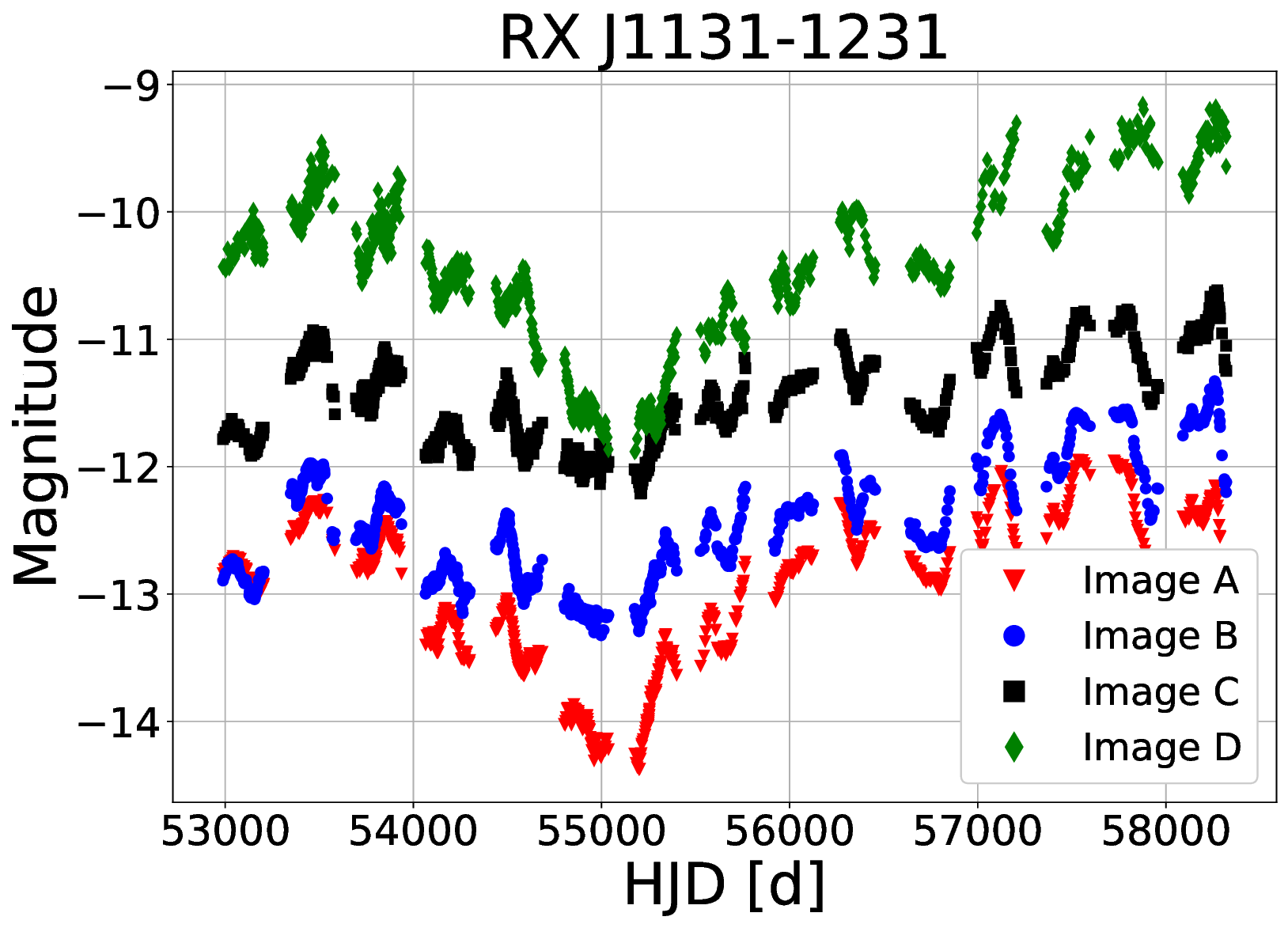}\par
\end{multicols}

\caption{Light Curves of the images of the present sample of lensed Quasars, obtained from the following references:   HE 0047-1756, UM 673, Q J0158-4325, HE 0435-1223, RX J1131-1231, SDSS J1226-0006, Q 1355-2257, SDSS J1620+1203, 2017WFI. J2026-4536, HE 2149-2745  \citep{millon2020cosmograil}, SDSSJ1004+4112 \citep{fohlmeister2007time}, SDSSJ1001+5027 \citep{kumar2013cosmograil}, SDSS J1226+4332 \citep{eulaers2013cosmograil}, and WFI 2033-4723 \citep{bonvin2019cosmograil}. Red, blue, black, and green colors correspond to the images A, B, C, and D, respectively.}
\label{fig:LC}
\end{figure*}

For the present study, we have taken into consideration a preliminary sample of 18 strongly lensed quasars given by \citet{millon2020cosmograil}, with photometric LCs covering about 15~years of monitoring, and four lensed quasars from \citet{fohlmeister2007time}, \citet{kumar2013cosmograil}, \citet{eulaers2013cosmograil}, and \citet{bonvin2019cosmograil}, providing us with a very long history in $r$ band observations, as shown in Fig.~\ref{fig:LC}. With this sample of 22 quasars in hand, we have imposed two criteria for the analysis as follows. (i) Only those objects presenting at least two different time series were considered because the lensing effect can produce different images of the same object. (ii) Only those LCs suitable for analysis with the WTMM method, which requires a minimum of about 350 data points to compute the multifractality parameters described in Section~\ref{sec:analysis}, were selected. Only 14 quasars met these criteria, which are listed in Table~\ref{tab:lens_properties} along with the redshift values for the lensing galaxy and the quasar source, $z_{lens}$ and $z_{source}$, respectively, as well as the time delay ($\Delta t_{AB}$). For these 14 lensed quasars, we have  taken additional information associated with microlensing from \citet{mosquera2011microlensing}, particularly those associated with the variability caused by the microlensing effect in the images, that is the characteristic size of the quasar source, $R_S$, the radius of the Einstein ring, $R_E$, and the characteristic size of the source’s broad line region, $R_{BLR}$. The Einstein ring was determined by the gravitational deflection induced by stars and compact objects within the lensing galaxy, using a mean mass of $<M>\,=0.3$~M$_{\odot}$ \citep{mosquera2011microlensing}. The characteristic timescales for microlensing variability, $t_S=R_S/v$ and $t_E=R_E/v$, where $v$ is the effective transverse velocity of the quasar source, were also obtained from \citet{mosquera2011microlensing}.
 Readers are referred to those authors for a complete description of the determination of the underlined parameters. In particular, the sizes of the accretion disks ($R_S$) listed in Table~\ref{tab:lens_properties} were estimated in the referred study from a simple thin-disk model \citet{shakura1973black}, scaled to the measured band flux. As estimated by \citet{morgan2010quasar}, $R_S$ uncertainties from band fluxes typically range within 0.1 to 0.2~dex on a logarithmic scale (about 30--60\%).

\subsection{Data analysis}
\label{sec:analysis}

There are several approaches to investigate the multifractality (nonlinearity) of time series, such as the Hurst analysis, also known as rescaled range analysis (R/S) (\citealt{hurst1951long}), and the Detrended Fluctuation Analysis (DFA) (\citealt{taqqu1995estimators}). An extension of the DFA is the Multifractal Detrended Fluctuation Analysis (MFDFA) (\citealt{kantelhardt2002multifractal}). Another possibility presented by \citet{gu2010detrending} uses Moving Averages (MA) for the multifractal analysis of time series and multifractal surfaces, namely Multifractal Detrended Moving Averages (MFDMA). The Wavelet Transform Modulus Maxima (WTMM) is also a powerful tool for the study of multifractality (\citealt{muzy1991wavelets,muzy1994multifractal}). Moreover, the WTMM can decompose the time and scale of the observed signal into fractal dimension regions, making it possible to identify the characteristics of multi-scale dimension in the time series. This method consists of two main steps: the wavelet analysis of the time series (in the present case, the lensed quasar LCs) and the multifractal formalism. To search for multifractality signature in the lensed quasars, we follow the same procedure used by \citet{belete2019revealing}, based on the WTMM transform. This method is discussed in a detailed description by \citet{belete2019revealing}  and the computational tools were adapted from \citet{puckovs2012wavelet}.

In summary, we have applied the procedure used by \citet{belete2019revealing} for each image of the analyzed lensed quasars as follows. First, we  computed the continuous wavelet transform $W(s, a)$ of each time series $x(t)$ from the formula \citep[e.g.,][]{addison2002illustrated}
	\begin{equation}
		 W(s, a) = \frac{1}{\sqrt{s}} \int_{0}^{T} \Psi \left( \frac{t - a}{s} \right) x(t) dt,
	\end{equation}
where \( T \) is the  time span of the series and $\Psi\left( \frac{t - a}{s} \right) $ denotes the wavelet mother function, namely the basic waveform to be dilated and translated according to different values of the scaling and shift parameters, $s$ and $a$, respectively. This integral expression characterizes the wavelet transform as a tool for analyzing signals, where the mother wavelet function is applied at varying scales and positions to capture signal features. For our case, we use  the Mexican hat as a mother function,
	\begin{equation}
		\Psi(t) = (1 - t^2) \cdot e ^{\frac{t^2}{2}},
	\end{equation}
from which information on their temporal variation is obtained for each scale factor chosen. This analysis  depends on the LC  length and  requires adjustments in the scale parameter  for each light curve. Numerically, the wavelet transform is represented by the wavelet coefficient matrix, $W_{s,a}$, which assumes a regularly sampled time series \citep{puckovs2012wavelet}. To handle data gaps in the time series, we interpolate the data to produce evenly distributed points, a process expected to minimally affect the results \citep{belete2018multifractality}. A more detailed discussion of the potential effects of these gaps, along with a quantified uncertainty analysis, is provided in Sect.~\ref{sec:uncertainties}.
The WTMM matrix provides the values of $W_{s,a}^{\rm abs}$ along its local maxima regions, where $W_{s,a}^{\rm abs}$ denotes the wavelet coefficient matrix with all elements expressed as absolute values. This is derived from the element-wise multiplication $WTMM = W_{s,a}^{\rm abs} \cdot LcMx_{s,a}$, where $LcMx_{s,a}$ is a boolean mask that identifies the local maxima regions of the $W_{s,a}^{\rm abs}$ matrix, typically outlining a fractal tree structure \citep{puckovs2012wavelet}.

From the WTMM matrix, we proceed with the following steps. (\romannum{1}) First, we compute \(Z_q(s)\), the thermodynamic partition function, connecting wavelet transform and multifractality analysis, given by \citet{belete2019revealing}, that is
\begin{equation}
  Z_q(s) = \sum_{a=1}^{T-1} [C(s) \cdot WTMM]^q | LcMx_{s,a} = 1],
\end{equation}
where $WTMM$ are the wavelet modulus maxima coefficients, $C(s)$ denotes a constant specific to each scale parameter, and \(q\) is the statistical moment, defined within a range of \([-5, 5]\). Moments of order \( q \) represent statistical characteristics computed as the expected value of differences between data points and their mean, raised to the power of \( q \), as given by
	\begin{equation}
		Z_q(s) \sim s^{\tau(q)}.
	\end{equation}

\begin{figure*}
    \centering
\begin{multicols}{2}
             {\includegraphics[width=\linewidth]{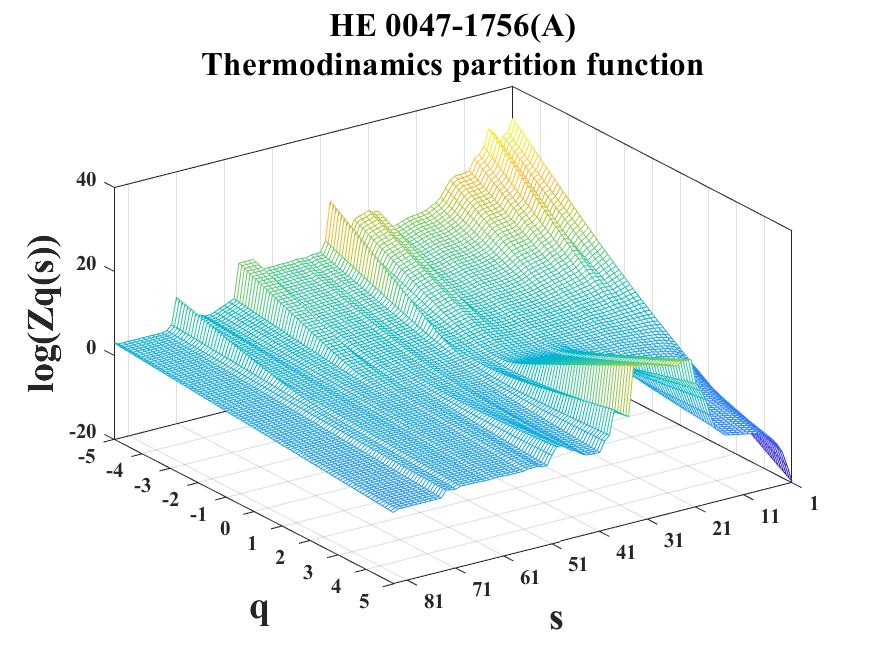}}\par
             {\includegraphics[width=\linewidth]{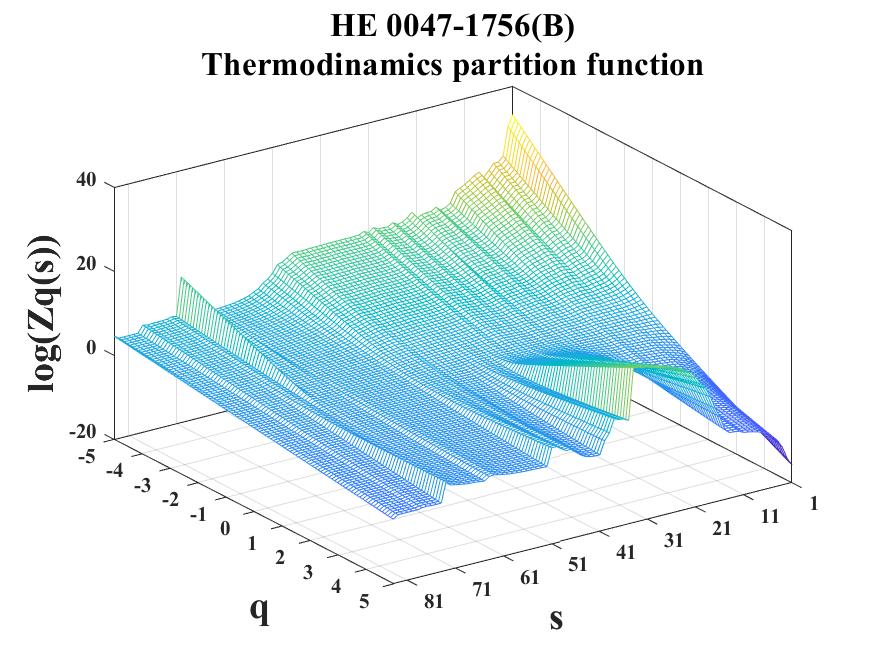}}\par
\end{multicols}
    \caption{ Thermodynamic partition function $Z_q(s)$ of images A (left panel) and B (right panel) of HE 0047-1756 in the $r$ band. The scale parameter $s$ is given in days, whereas the statistical moment, $q$, and partition function, $Z_q(s)$, are dimensionless quantities.
    }
    \label{fig:WTMM_sep_HE 0047-1756_example}
\end{figure*}

\begin{figure*}
    \centering
\begin{multicols}{2}            {\includegraphics[width=\linewidth]{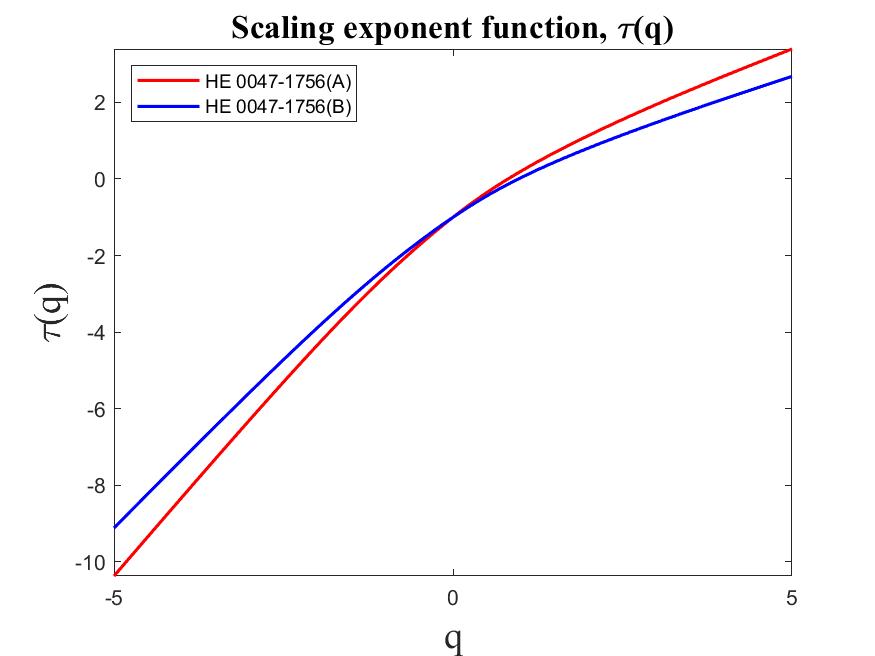}}\par
{\includegraphics[width=\linewidth]{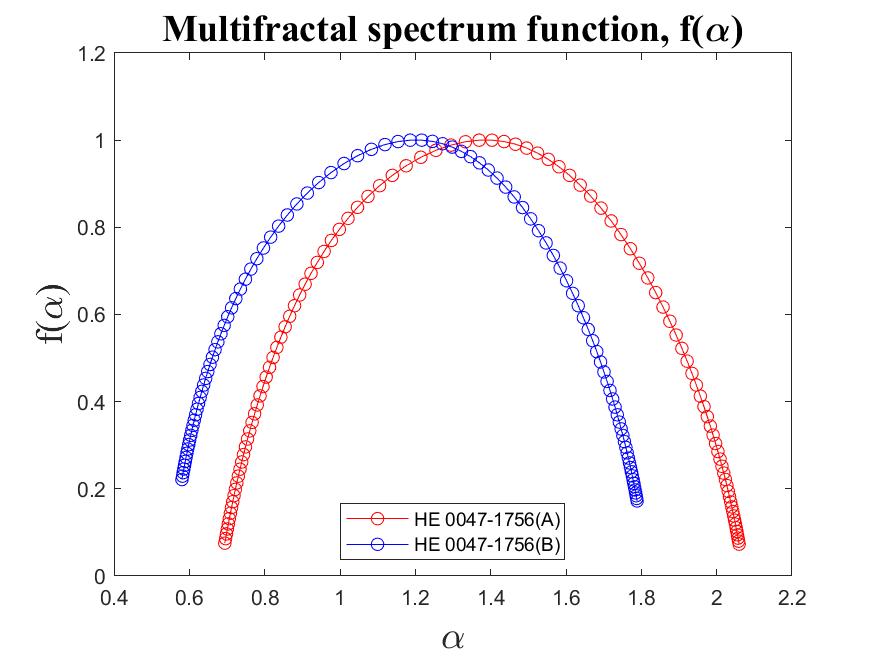}}\par
\end{multicols}
    \caption{The scaling exponent functions $\tau (q)$ (left panel) and the multifractal spectrum functions $f(\alpha)$ (right panel) for image A (red) and image B (blue) of HE 0047-1756 in the $r$ band.}
    \label{fig:WTMM_HE 0047-1756_example}
\end{figure*}

The relation of the partition function to the scale parameter, $s$, is presented as an example in  Figs.~\ref{fig:WTMM_sep_HE 0047-1756_example} and~\ref{fig:WTMM_HE 0047-1756_example}, for the lensed quasar HE 0047-1756, where there is a nonlinear behavior between the scale parameter and the partition function. (\romannum{2}) The $Z_q(s)$ fluctuations were then used to determine the scale exponent function, $\tau(q)$, where $q$ is the  statistical moment. The relationship between the scaling exponent function and  this moment is also nonlinear, as shown in the middle column of Fig.~\ref{fig:WTMM_sep_HE 0047-1756_example}, which indicates the presence of a multifractal (nonlinear) behavior of the time series (i.e., the difference of the curve shown from a straight line represents nonlinearity). (\romannum{3}) We then calculated the degree of multifractality (nonlinearity) present in each LC. To determine the degree of multifractality, we calculated the multifractal spectrum function, $f(\alpha )$: 
\begin{equation}
	f(\alpha) = q\cdot \alpha - \tau(q),
	\label{f_alpha}
\end{equation}
which is dependent on the Holder's exponent, $\alpha$ ~(\citealt{halsey1986fractal}):
\begin{equation}
	\alpha = \alpha(q) = \frac{\partial \tau(q)}{\partial q}.
	\label{Holder}
\end{equation}
 For instance, $f(\alpha)$ typically has a bump shape with a specific width that depends on the characteristics of each time series (e.g., Belete et al. 2018, MNRAS, 478, 3976). For a given $f(\alpha)$, the value of $\Delta\alpha$ indicates the level of multifractality in the sense that smaller values of $\Delta\alpha$ (i.e., $\Delta\alpha$ near zero) indicate the monofractal limit. In contrast, larger values indicate the strength of the multifractal behaviour in the signal \citep{shimizu2002multifractal,ashkenazy2003nonlinearity,telesca2004investigating}.

Next, we calculated the degree of multifractality using the Eqs. (\ref{f_alpha}) and (\ref{Holder}) for each image, as the example shown in the right column of Fig.~\ref{fig:WTMM_sep_HE 0047-1756_example}. Finally, the width $\Delta \alpha = \alpha_{max} - \alpha_{min}$ was computed for the multifractal spectrum of each image analyzed. Those widths are shown in Table~\ref{tab:multifractalidade}.

\section{Results}
\label{sec:Results}

This section presents the results of our multifractal analysis of the time series of  the images of 14 gravitational lensed quasars in the redshift range of 0.657--2.730. For a better presentation of the results obtained with the WTMM analysis, we used the number of images ($N$) formed by the lensing effect of each quasar, separating the systems into two groups: one with quasars containing two images ($N=2$), which allows a direct comparison with the finding of \citet{belete2019revealing} for the quasar Q0957+561, and another group with quasars containing three and four images ($N>2$). Indeed, the number of formed images is related to the relative  positions between the lensing galaxy, the quasar source, and the observer, as well as to the gravitational potential of the  lensing galaxy.

The analysis was  conducted following the same approach applied by \citet{belete2019revealing}, first with the computation of the thermodynamic partition function, the scaling exponent, and the multifractal spectrum, as described in Section \ref{sec:observations}. These results are displayed in Fig.~\ref{fig:WTMM_sep_HE 0047-1756_example}, where the slope of $\log Z_q (s)$ as a function of $\log (s)$, represented by $\tau (q)$, clearly indicates the presence of multifractal structures in all the LCs of the present sample of lensed quasars, a result confirmed by the behavior of the exponent functions of scale $\tau (q)$ (see \citealp{belete2019revealing}). Figure~\ref{fig:WTMM_sep_HE 0047-1756_example} also shows the multifractal spectrum functions $f(\alpha)$, where the width $\Delta \alpha$ further confirms the differences in the intensity of the nonlinearity between the different images.

The degree of the multifractality for each image was computed from the width $\Delta \alpha$ of the multifractal spectrum of the corresponding image. The obtained values of the degree of multifractality for each system are listed in Table~\ref{tab:multifractalidade}, where $\Delta \alpha_A$, $\Delta \alpha_B$, $\Delta \alpha_C$, and $\Delta \alpha_D$ represent the width of the multifractal spectrum functions for images A, B, C, and D, respectively. For all the systems with two images, one observes different values for the degree of multifractality, $\Delta \alpha_A$ and $\Delta \alpha_B$, namely $\delta_{AB} \neq 1 $, indicating that images of the same quasar have distinct multifractals characteristics, corroborating the finding by \citet{belete2019revealing}. The systems with $N > 2$, namely those with 3 and 4 images, also present different values for the degree of multifractality, $\Delta \alpha_A$, $\Delta \alpha_B$, $\Delta \alpha_C$, and $\Delta \alpha_D$, following the scenario for the multifractality degree observed for quasar systems with only two images.

As a second step, we calculated the  ratio between the multifractal lengths of each image,  given by 
\begin{equation}
	\delta_{ij} =  \dfrac{\Delta\alpha_{i}}{\Delta\alpha_{j}}.
	\label{eq:Excess_multifractal}
\end{equation}
Here, $\delta_{ij}$ represents the excess of multifractality between a quasar image $i$ compared to image $j$, where indices $i$ and $j$ denote images A, B, C, or D.
The parameter $\delta_{ij}$ can inform how the processes that affect the formation of images $i$ and $j$  can be different internally and externally. The significance of this magnitude can be discerned in Fig.~\ref{fig:WTMM_HE 0047-1756_example}, where clear differences in the degrees of nonlinearity are evident among the images, both in the behavior of the scaling factor (on the left) and in the computation of the multifractal spectrum (on the right), exhibiting notable distinctions. Following the same criteria used by \citet{belete2019revealing}, $\delta_{ij}$ is greater than 1 when the multifractality of image $j$ is higher than that of image $i$, and with $\delta_{ij} < 1$ in the opposite case. These scenarios indicate that at least one of the variables of one of the images is being affected by external factors to the quasar source. When $\delta_{AB} = 1$, there is no excess, and it indicates that internal processes are predominant or unique for the presence of nonlinearity. As pointed out by~\citet{belete2019revealing}, the light curves from different images likely exhibit similar behaviors, except for some lags and an overall magnitude offset \citep{wambsganss1998gravitational}. Thus, in the absence of extrinsic variations, such as microlensing, their non-linear signatures would remain similar. Therefore, assuming that the $r$-band signals have comparable radiation mechanisms (from the accretion disc or a compact source), any differences in their non-linearity strength would likely stem from extrinsic variabilities or microlensing effects by stars in the lensing galaxies~\citep{kostrzewa2018gravitationally}.

The excess of the degree of multifractality between the different images of the systems,  $\delta_{AB}$, $\delta_{AC}$, $\delta_{AD}$, $\delta_{BC}$, $\delta_{BD}$, and $\delta_{CD}$, was then computed and is listed in Table~\ref{tab:mexcess_multifractal}. Indeed, for systems with two images (A and B), the $\delta_{AB}$ values show the multifractality difference between images A and B; for the systems with three images (A, B, and C), the values of $\delta_{AC}$ and $\delta_{BC}$ represent the differences in multifractality between images A and C, and B and C, respectively; for the systems with four images (A, B, C, and D), the values of $\delta_{AD}$, $\delta_{BD}$, and $\delta_{CD}$ indicate the differences in multifractality between all possible combinations of images.
These differences may provide valuable information about the intrinsic nature of the LCs in the analyzed lens systems, highlighting distinct multifractal features in each image associated with different physical processes or conditions of the intergalactic environment.

 According to \citet{belete2019revealing}, the degree of multifractality of different images of the same quasar can vary due to potential influences from intrinsic and extrinsic factors. Intrinsic factors include variations in flux from the accretion disk or continuum compact source, which should theoretically yield similar multifractal behaviors across images unless there are specific time delays or offsets in magnitude \citep{wambsganss1998gravitational}. Extrinsic factors, such as microlensing effects from stars in the lensing galaxies, can induce differences in multifractality between images by affecting the observed light curves \citep{kostrzewa2018gravitationally}. Therefore, while the underlying radiation mechanisms and regions may be similar across images, external influences and observational conditions can result in different degrees of multifractality.

\subsection{Uncertainties and possible biases on the multifractality calculation}
\label{sec:uncertainties} 

\begin{figure}
    \centering
\includegraphics[width=\columnwidth]{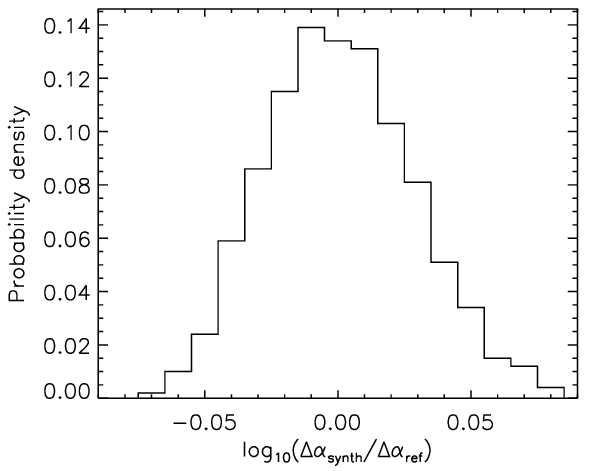}
    \caption{Distribution of the deviations of $\Delta \alpha_A$ and $\Delta \alpha_B$, altogether, labeled $\Delta \alpha_{\rm synth}$, computed from a set of 1000 synthetic light curves simulating the effects caused by time delays, compared to their expected values, $\Delta \alpha_{\rm ref}$, as explained in the text.
    }
    \label{fig:mock_stats}
\end{figure}

We tested the stability of the WTMM method by applying it to synthetic light curves. As a first test, we investigated the observational effects on lensed images with different time delays \citep{tewes2013cosmgrail9}. Although the time delay is essentially a temporal translation, a larger time delay results in a larger difference in the path traversed by the deflected light. Consequently, the lensed images formed in different regions may present distinct variations in observed brightness, due to the different trajectories of light around the lens \citep{hawkins2020signature}. As such, we employed a dataset comprising 1000 synthetic light curves of the quadruple system HE0435-1223, as provided by \citet{bonvin2019cosmograil}, with random time delays within $\pm$6~days. This allowed us to examine how these physical effects might influence the degree of multifractality calculated using the WTMM method. For a global view of this analysis, Fig.~\ref{fig:mock_stats} displays the statistical distribution of the deviation of $\Delta \alpha_A$ and $\Delta \alpha_B$, altogether, named $\Delta \alpha_{\rm synth}$, computed from synthetic data, with respect to their expected values, $\Delta \alpha_{\rm ref}$. The latter represents to the most probable values for $\Delta \alpha_A$ and $\Delta \alpha_B$ for the HE0435-1223 light curves, based on the synthetic data, corresponding to no time delay. The standard deviation of this distribution is approximately 7\% ($\sim$0.03~dex on a logarithmic scale), providing an order of magnitude of the uncertainties associated with the influence of the time delays on the degree of multifractality.

\begin{figure}
    \centering
\includegraphics[width=\columnwidth]{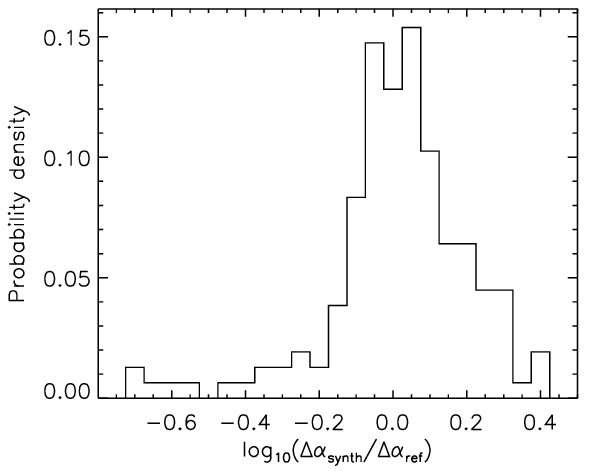}
    \caption{Distribution of the deviations of $\Delta\alpha_A$ and $\Delta\alpha_B$, altogether, labeled $\Delta \alpha_{\rm synth}$, computed from synthetic light curves with random gaps, compared to the same parameters computed from the original light curves, $\Delta \alpha_{\rm ref}$. The original light curves are those from the sample of 14 lensed quasars considered in this work.
    }
    \label{fig:hist_gaps}
\end{figure}

In addition, we analyzed how $\Delta\alpha$ could be affected by observational limitations, such as the presence of gaps in the light curves or the lengths of the time series. To test the influence of gaps, we injected synthetic gaps into the light curves of the 14 quasars considered in this work. The gaps were randomly distributed and with different durations, based on the statistics of the actual gaps, which typically correspond to about 20--30\% of missing data.
This test showed that introducing random gaps caused $\Delta\alpha$ to vary within a standard deviation of $\sim$30--60\% ($\sim$0.1--0.2~dex on a logarithmic scale). Figure~\ref{fig:hist_gaps} illustrates a global view of this analysis. Furthermore, we also investigated the influence of the lengths of the light curves, and no clear correlation between $\Delta \alpha$ and the time span was observed. Nevertheless, different lengths do produce a fluctuation in the $\Delta \alpha$ values, similar to the presence of gaps.

\begin{figure}
    \centering
\includegraphics[width=\columnwidth]{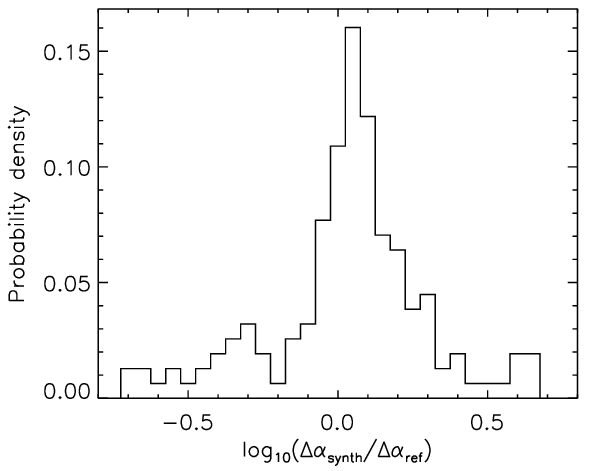}
    \caption{Distribution of the deviations of $\Delta\alpha_A$ and $\Delta\alpha_B$, altogether, labeled $\Delta \alpha_{\rm synth}$, computed from synthetic light curves with different levels of gaussian noise, compared to the same parameters computed from the original light curves, $\Delta \alpha_{\rm ref}$. The original light curves are those from the sample of 14 lensed quasars considered in this work.}
    \label{fig:hist_noises}
\end{figure}

Finally, we tested the influence of high-frequency noises on the multifractality measurement by smoothing the light curves of Fig.~\ref{fig:LC} with a boxcar average of typically 20~days and using them as illustrative noise-free models. The residual of the original light curves detrended from those models resulted in guassian noises with typicaly 10\% of the main signal amplitudes. Then, we tested adding gaussian noises of different levels, specifically from about 1\% to 20\% of the signal amplitudes, on the noise-free models to study how those noises affect the $\Delta\alpha$ values. Figure~\ref{fig:hist_noises} displays the distribution of the deviations of synthetic $\Delta\alpha$ values, obtained from the light curve models with different gaussian noises added, with respect to their corresponding values from the original light curves. The typical deviation of those tests also lies around 0.1--0.2~dex.

Overall, based on all performed tests, we suggest a typical uncertainty of $\sim$0.1--0.2~dex for the $\Delta\alpha$ values computed in this study, associated with observational limitations. This uncertainty is dominant over the one related to time delays estimated above and can be considered a global uncertainty for $\Delta\alpha$ based on the present analysis.

\subsection{Is there a connection between multifractality, sizes and timescales of lensed quasars?}
\label{sec:relation} 

Despite the possible biases associated with the limited sample of quasar systems used in this study, it is worthwhile to investigate potential relationships between the degree of multifractality exhibited by images A and B, represented by $\Delta\alpha_A$ and $\Delta\alpha_B$, respectively, and the characteristic parameters related to the quasar, as well as the quasar source. These parameters include the Einstein ring ($R_E$) and the accretion disk size ($R_S$), as well as the characteristic timescales related to microlensing variability. Let us underline that, in the present quasar sample, the timescales of microlensing variability range from a few months to a few years \citep[e.g.,][]{stone2022optical}, thus compatible with the time span of the light curves considered in the present study, ranging within about~2--15~years. Indeed, microlensing effects have been identified in several objects analyzed here using different methods~\citet{hawkins2010time,mosquera2011microlensing,hawkins2020sdss,hawkins2020signature,hawkins2022new}. The timescale $t_E$ is proportional to the distance traveled by the quasar source radiation, equivalent to one Einstein radius. At the same time, $t_S$ represents the time the light takes to cross the source size. Also, the amplitude of the variations in brightness observed in the multiple images of the quasar will be influenced by the ratio of the angular size of the quasar's source ($R_S$) to the Einstein radius ($R_E$) of the gravitational lensing object. Specifically, when the $R_S/R_E$ ratio is smaller, this leads to larger amplitudes of brightness variations in the lensed images, implying that the quasar's source is more compact relative to the size of the lensing object. Therefore, the magnification of the source is more sensitive to small changes in the alignment between the source and the lens. It is also relevant to analyze the role of microlensing on the degree of multifractality, an aspect well-explored by \citet{belete2019revealing}. According to the refereed work, microlensing has a clear influence on multifractality. Any discrepancy in the degree of multifractality between different images of a quasar is expected to be caused by extrinsic variabilities of different origins or due to microlensing by stars in the lensing galaxies affecting the images (Kostrzewa-Rutkowska et al. 2018).

\begin{figure}
    \centering
            {\includegraphics[width=\columnwidth]{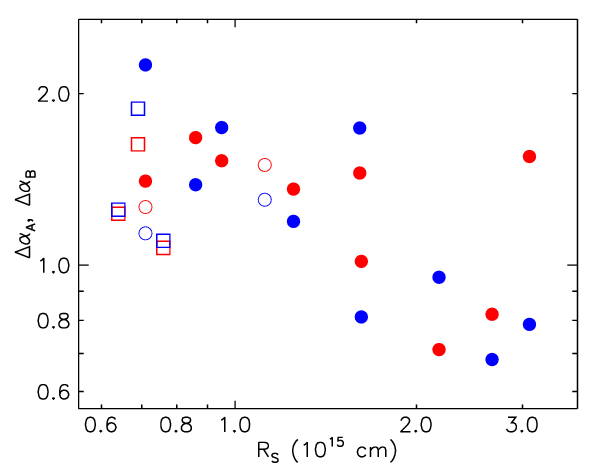}}\par
    \caption{
Distribution of the accretion disk size, $R_S$, as a function of the degree of multifractality, $\Delta \alpha$, for the present sample consisting of 14 lensed quasars.
Red and blue symbols correspond to quasar images A and B, respectively. Filled circles denote lensed quasars with two images, open circles those with three images, and open squares represent lensed quasars with four images.
}
    \label{fig:results}
\end{figure}

Fig.~\ref{fig:results} shows the behavior of the degree of multifractality ($\Delta \alpha$) as a function of the accretion disk size, $R_S$, from where one observes  a possible trend, with a decreasing of $R_S$ with the increasing of $\Delta \alpha$. 
Due to the typical uncertainties associated with these parameters (see Sections~\ref{sec:datacollection} and~\ref{sec:uncertainties}), it remains unclear whether this trend is physical or influenced by biases. It is important to notice that this concern arises from observational limitations rather than inherent to the method.
Considering that $R_S$ is related to the timescale, $t_S$, this possible outcome suggests that the degree of multifractality may be linked with the size or the timescale of the accretion disk. A statistical analysis of the $R_S$ versus $\Delta \alpha$ relation, based on the Spearman rank correlation, $\rho$, points for a solid correlation with $\rho_{\Delta \alpha_B} = -0.73$ for image~B. For image~A, considering the bulk of the data, we obtain $\rho_{\Delta \alpha_A} = -0.28$. Nevertheless, one object, HE 2149-2745, with $\Delta\alpha = 1.551$ and $R_S = 3.08 \times 10^{15}$~cm, presents a deviation from this trend. Such a discrepancy may be attributed to the complex internal structure of this object, identified as a Broad Absorption Line (BAL) quasar, as reported by \citet{millon2020cosmgrail}. Without this object, $\rho_{\Delta \alpha_A} = -0.65$, following closely the correlation observed for image~B. Therefore, the rank statistics support the possibility of a correlation between $R_S$ and $\Delta \alpha$. Considering the scope of our analysis, further data is necessary to establish a conclusive understanding of this aspect.



%
%

\section{Summary and Conclusions}
\label{sec:conclusions}

The detection of strong multifractal signatures in the LCs of the two images of the quasar Q0957+561 was first carried out by \citet{belete2019revealing}. The degree of multifractality for both images, changing over time in a non-monotonic way, was interpreted as revealing the presence of extrinsic variabilities in the LCs of the images. Here, we applied the same procedure used by those authors, now for an enlarged sample of 14 lensed quasars, nine with two images, two with three images, and three with four images, aiming to identify similar multifractal signatures. In short, first, we computed the absolute wavelet coefficients using the continuous wavelet transform approach, and, using the constructed skeleton function, we determined the thermodynamics partition function for the LCs of all the considered quasar systems. Second, we estimated the slope of the log–log plots of the thermodynamic partition function $Z_q (s)$ and the scale $s$, quantified by the scaling exponent function $\tau (q)$ versus the moment $q$ plots. Finally, we estimated the multifractal spectrum at each frequency for all the LCs and computed the degree of multifractality from the width $\Delta \alpha$.

The first relevant scenario emerging from the present study concerns the identification of multifractality signatures in the LCs of the images of the analyzed lensed quasar systems, confirming the finding by \citet{belete2019revealing}, for quasar Q0957+561. Such an aspect is observed for all the analyzed quasar systems, independently of the number of images, with a significant difference between the degree of multifractality of all the images A, B, C, and D, and combinations. As pointed out by \citet{belete2019revealing}, a difference between the degree of multifractality tending to one indicates that internal processes are predominant or unique for the presence of nonlinearity. In contrast, a difference between the degree of multifractality different from one indicates that at least one of the variables of one of the images is being affected by factors external to the quasar source. Despite the presence of multifractality signatures in all the LC images, there is no clear relation between the strength of the degree of multifractality of one LC image once compared with its pair. Indeed, for nine quasars, the degree of multifractality of image A is greater than that of B, whereas for six quasars, there is an opposite scenario. 

We have also searched for a possible connection between the degree of multifractality $\Delta\alpha$ and the accretion disk size ($R_S$). This analysis reveals some apparent trends with a decrease of $R_S$ with the increase of $\Delta\alpha$, pointing to a decrease in the degree of multifractality with the increase of the source size and timescale. This result suggests that the complexity level, as measured in our analysis, depends on internal factors of the quasar source. Based on our analyses, the constant differences in multifractality between images A and B of the quasars remain uncertain. Nevertheless, we should be cautious with the referred apparent trend because different properties of the LCs, like the signal-to-noise or temporal variability of their amplitudes and potential bias associated with the sample limitation, may impact the observed behavior.


\begin{table*}
    \centering
    \caption{Summary of the lens properties. The redshift ($z$) and time delay ($\Delta t$) values were obtained from: HE 0047-1756, UM 673, Q J0158-4325, HE 0435-1223, RX J1131-1231, SDSS J1226-0006, Q 1355-2257, SDSS J1620+1203, WFI J2026-4536, HE 2149-2745 \citep{millon2020cosmograil}, SDSSJ1004+4112 \citep{fohlmeister2007time}, SDSSJ1001+5027 \citep{kumar2013cosmograil}, SDSS J1226+4332 \citep{eulaers2013cosmograil}, and WFI 2033-4723 \citep{bonvin2019cosmograil}. Values of the Einstein radius ($R_E$) and quasar's source radius ($R_S$) were obtained from \citet{mosquera2011microlensing}.}
    \label{tab:lens_properties}
    \begin{tabular}{cccccccc} 
        \hline
        \hline
        Lens & $z_{lens}$ & $z_{source}$ & $m_s$ & $R_E (10^{16} \mathrm{cm})$ & $R_S  $ $(10^{15} \mathrm{cm})$ & $\Delta t_{AB}$ (days) & N \\
        \hline
        HE 0047-1756 & 0.407 & 1.678 & 16.52 & 3.12 & 1.25 & -10.4 & 2 \\
        HE 2149-2745 & 0.603 & 2.033 & 16.29 & 2.86 & 3.08 & -39 & 2 \\
        UM 673 & 0.491 & 2.73 & 16.47 & 2.84 & 2.67 & -97.7 & 2 \\
        QJ0158-4325 & 0.317 & 1.29 & 17.39 & 3.41 & 1.62 & -22.7 & 2 \\
        Q 1355-2257 & 0.039 & 1.69 & 16.94 & 2.92 & 2.18 & -81.5 & 2 \\
        SDSS J1226-0006 & 0.517 & 1.123 & 18.3 & 2.35 & 0.86 & 33.7 & 2 \\
        SDSS J1001+5027 & 0.84 & 1.84 & 17.31 & 3.08 & 1.61 & -119.3 & 2 \\
        SDSS J1206+4332 & 0.748 & 1.79 & 18.47 & 3.11 & 0.71 & -111.3 & 2 \\
        SDSS J1620+1203 & 0.398 & 1.158 & 19.1 & 2.87 & 0.95 & -171.5 & 2 \\
        WFI J2026-4536 & $\sim$ 1.040 & 2.23 & 16.18 & 2.13 & 1.12 & 18.7 & 3 \\
        WFI 2033-4723 & 0.66 & 1.66 & 17.59 & 2.37 & 0.71 & 36.2 & 3 \\
        SDSSJ1004+4112 & 0.68 & 1.73 & 17.53 & 2.35 & 0.69 & -38.4 & 4 \\
        HE 0435-1223 & 0.454 & 1.693 & 16.84 & 2.94 & 0.76 & -9 & 4 \\
        RX J1131-1231 & 0.295 & 0.657 & 16.74 & 2.5 & 0.64 & 1.6 & 4 \\
        \hline
    \end{tabular}
\end{table*}

 \begin{table*}
	\centering
	\caption{Multifractal analysis for the present sample of lensed quasars. Columns labeled $\Delta \alpha_{A}$, $\Delta \alpha_{B}$, $\Delta \alpha_{C}$, and $\Delta \alpha_{D}$ refer to the degrees of multifractality computed in this work for the respective images, A, B, C, and D.}
        \label{tab:multifractalidade}
	\begin{tabular}{ccccc} 
		\hline
		\hline
		Lens & $\Delta \alpha _{A}$ & $\Delta \alpha _{B}$ & $\Delta \alpha _{C}$  & $\Delta \alpha _{D}$ \\
		\hline
		HE 0047-1756      & 1.365 & 1.208   & -     & -  \\
		HE 2149-2745      & 1.551 & 0.787   & -     & -  \\
		UM 673            & 0.819 & 0.802   & -     & -  \\
		QJ0158-4325       & 1.015 & 0.811   & -     & -  \\
		Q 1355-2257       & 0.710 & 0.952   & -     & -  \\
		SDSS J1226-0006   & 1.674 & 1.383   & -     & -  \\
		SDSS J1001+5027	  & 1.865 & 1.740   & -     & -  \\
		SDSS J1206+4332	  & 1.404 & 2.245   & -     & -  \\
		SDSS J1620+1203   & 1.524 & 1.743   & -     & -  \\
            WFI J2026-4536    & 1.499 & 1.302   & 1.055 & -  \\
		2017WFI 2033-4723 & 1.264 & 1.137   & 1.072 & -  \\
            SDSSJ1004+4112    & 1.600 &	1.939   & 1.054 & 1.292 \\
		HE 0435-1223      & 1.072 & 1.103   & 1.180 & 1.075 \\
		RX J1131-1231     & 1.230 & 1.251   & 1.334 & 0.902  \\
		\hline
	\end{tabular}
   \end{table*}

\begin{table*}
	\centering
	\caption{Multifractality ($\delta_{ij}$) for the analyzed quasars. $\delta_{ij}$ represents the ratio between the degrees of multifractality ($\Delta \alpha_{i}$ and $\Delta \alpha_{j}$) for different image pairs.}
        \label{tab:mexcess_multifractal}
\begin{tabular}{ccccccc} 
	\hline
	\hline
  Lens & $\delta_{AB}$ & $\delta_{AC}$ & $\delta_{AD}$ & $\delta_{BC}$ & $\delta_{BD}$ & $\delta_{CD}$ \\
		\hline
	HE 0047-1756      & 1.130 & - & - & - & - & -\\
	HE 2149-2745      & 1.974 & - & - & - & - & -\\
	UM 673            & 1.022 & - & - & - & - & -\\
	QJ0158-4325       & 1.252 & - & - & - & - & -\\
	Q 1355-2257       & 0.746 & - & - & - & - & -\\
	SDSS J1226-0006   & 1.211 & - & - & - & - & -\\
	SDSS J1001+5027	  & 1.071 & - & - & - & - & -\\
	SDSS J1206+4332	  & 0.625 & - & - & - & - & -\\
	SDSS J1620+1203   & 0.876 & - & - & - & - & -\\
	WFI J2026-4536    & 1.150 & 1.144 & - & 1.073 & - & -\\
	2017WFI 2033-4723 & 1.112 & 1.180 & - & 1.080 & - & -\\
	SDSSJ1004+4112    & 0.825 & 0.825 & 1.485 & 0.789 & 1.510 & 1.821 \\
	HE 0435-1223      & 0.972 & 0.910 & 0.908 & 0.932 & 1.037 & 0.928 \\
	RX J1131-1231     & 0.982 & 0.922 & 1.004 & 0.985 & 1.385 & 1.538 \\
	\hline
\end{tabular}
\end{table*}

\section*{Acknowledgements}

Research activities of the observational astronomy board at the Federal University of Rio Grande do Norte are supported by continuous grants from the Brazilian funding agencies CNPq, FAPERN, and INCT-INEspaço. This study was financed in part by the Coordenação de Aperfeiçoamento de Pessoal de Nível Superior - Brasil (CAPES) - Finance Code 001. RAA and JPSC acknowledge CAPES graduate fellowships and LMCA acknowledges CNPq/PIBIC undergraduate fellowship. ICL, BLCM, and JRM acknowledge CNPq research fellowships. This research has made use of the VizieR catalogue access tool, CDS,
Strasbourg, France (DOI :  \href{https://doi.org/10.26093/cds/vizier}{10.26093/cds/vizier}). The original description of the VizieR service was published in \citet{vizier2000}. We warmly thank the Referee for comments and suggestions that clarified important aspects of this study.

\section*{Data Availability}

Light curves are available at the CDS via anonymous ftp to cdsarc.u-strasbg.fr (130.79.128.5) or via \url{http://cdsarc.u-strasbg.fr/viz-bin/cat/J/A+A/640/A105} for HE 0047-1756, UM 673, Q J0158-4325, HE 0435-1223, RX J1131-1231, SDSS J1226-0006, Q 1355-2257, SDSS J1620+1203, WFI J2026-4536, HE 2149-2745; via \url{http://cdsarc.u-strasbg.fr/viz-bin/qcat?J/A+A/557/A44} for SDSS J1001+5027; via \url{http://cdsarc.u-strasbg.fr/viz-bin/qcat?J/A+A/553/A121} for SDSS J1226+4332, and via \url{http://cdsarc.u-strasbg.fr/viz-bin/cat/J/A+A/629/A97} for WFI J2033-4723. For SDSS J1004+4112, light curves are available at the NASA/IPAC Extragalactic Database. For the mock light curves of HE0435-1223, \url{https://shsuyu.github.io/H0LiCOW/site/h0licow_data.html}4357. All data generated or analyzed during this study are included in this published article.


\bibliographystyle{mnras}
\bibliography{bibliografia} 

\begin{thebibliography}{}
\makeatletter
\relax
\def\mn@urlcharsother{\let\do\@makeother \do\$\do\&\do\#\do\^\do\_\do\%\do\~}
\def\mn@doi{\begingroup\mn@urlcharsother \@ifnextchar [ {\mn@doi@}
  {\mn@doi@[]}}
\def\mn@doi@[#1]#2{\def\@tempa{#1}\ifx\@tempa\@empty \href
  {http://dx.doi.org/#2} {doi:#2}\else \href {http://dx.doi.org/#2} {#1}\fi
  \endgroup}
\def\mn@eprint#1#2{\mn@eprint@#1:#2::\@nil}
\def\mn@eprint@arXiv#1{\href {http://arxiv.org/abs/#1} {{\tt arXiv:#1}}}
\def\mn@eprint@dblp#1{\href {http://dblp.uni-trier.de/rec/bibtex/#1.xml}
  {dblp:#1}}
\def\mn@eprint@#1:#2:#3:#4\@nil{\def\@tempa {#1}\def\@tempb {#2}\def\@tempc
  {#3}\ifx \@tempc \@empty \let \@tempc \@tempb \let \@tempb \@tempa \fi \ifx
  \@tempb \@empty \def\@tempb {arXiv}\fi \@ifundefined
  {mn@eprint@\@tempb}{\@tempb:\@tempc}{\expandafter \expandafter \csname
  mn@eprint@\@tempb\endcsname \expandafter{\@tempc}}}

\bibitem[\protect\citeauthoryear{Abajas, Mediavilla, Mu{\~n}oz, Popovi{\'c}  \&
  Oscoz}{Abajas et~al.}{2002}]{abajas2002influence}
Abajas C.,  Mediavilla E.,  Mu{\~n}oz J.,  Popovi{\'c} L.,   Oscoz A.,  2002,
  The Astrophysical Journal, 576, 640

\bibitem[\protect\citeauthoryear{Addison}{Addison}{2002}]{addison2002illustrated}
Addison P.,  2002, The Illustrated Wavelet Transform Handbook: Introductory
  Theory and Applications in Science, Engineering, Medicine and Finance.
Taylor \& Francis, \url {https://books.google.com.br/books?id=RUSjIMQACQQC}

\bibitem[\protect\citeauthoryear{Ashkenazy, Baker, Gildor  \& Havlin}{Ashkenazy
  et~al.}{2003}]{ashkenazy2003nonlinearity}
Ashkenazy Y.,  Baker D.~R.,  Gildor H.,   Havlin S.,  2003, Geophysical
  research letters, 30

\bibitem[\protect\citeauthoryear{Barr, Bremer, Baker  \& Lehnert}{Barr
  et~al.}{2003}]{barr2003cluster}
Barr J.,  Bremer M.,  Baker J.,   Lehnert M.,  2003, Monthly Notices of the
  Royal Astronomical Society, 346, 229

\bibitem[\protect\citeauthoryear{Bashan, Bartsch, Kantelhardt  \&
  Havlin}{Bashan et~al.}{2008}]{bashan2008comparison}
Bashan A.,  Bartsch R.,  Kantelhardt J.~W.,   Havlin S.,  2008, Physica A:
  Statistical Mechanics and its Applications, 387, 5080

\bibitem[\protect\citeauthoryear{Belete, Bravo, Canto~Martins, Leao, De~Araujo
  \& De~Medeiros}{Belete et~al.}{2018}]{belete2018multifractality}
Belete A.~B.,  Bravo J.,  Canto~Martins B.,  Leao I.,  De~Araujo J.,
  De~Medeiros J.,  2018, Monthly Notices of the Royal Astronomical Society,
  478, 3976

\bibitem[\protect\citeauthoryear{Belete, Canto~Martins, Le{\~a}o  \&
  De~Medeiros}{Belete et~al.}{2019a}]{belete2019revealing}
Belete A.~B.,  Canto~Martins B.,  Le{\~a}o I.,   De~Medeiros J.,  2019a,
  Monthly Notices of the Royal Astronomical Society, 484, 3552

\bibitem[\protect\citeauthoryear{Belete, Femmam, Tornikosk,
  L{\"a}hteenm{\"a}ki, Tammi, Le{\~a}o, Canto~Martins  \& De~Medeiros}{Belete
  et~al.}{2019b}]{belete2019cosmological}
Belete A.~B.,  Femmam S.,  Tornikosk M.,  L{\"a}hteenm{\"a}ki A.,  Tammi J.,
  Le{\~a}o I.,  Canto~Martins B.,   De~Medeiros J.,  2019b, The Astrophysical
  Journal, 873, 108

\bibitem[\protect\citeauthoryear{Belete, Goicoechea, Leao, Martins  \&
  De~Medeiros}{Belete et~al.}{2019c}]{belete2019novel}
Belete A.~B.,  Goicoechea L.,  Leao I.,  Martins B.~C.,   De~Medeiros J.,
  2019c, The Astrophysical Journal, 879, 113

\bibitem[\protect\citeauthoryear{Belete, Goicoechea, Canto~Martins, Le{\~a}o
  \& De~Medeiros}{Belete et~al.}{2020}]{bewketu2020nature}
Belete A.~B.,  Goicoechea L.,  Canto~Martins B.,  Le{\~a}o I.,   De~Medeiros
  J.,  2020, Monthly Notices of the Royal Astronomical Society, 496, 784

\bibitem[\protect\citeauthoryear{Belete et~al.,}{Belete
  et~al.}{2021}]{belete2021molecular}
Belete A.~B.,  et~al., 2021, Astronomy \& Astrophysics, 654, A24

\bibitem[\protect\citeauthoryear{Bonvin, Tewes, Courbin, Kuntzer, Sluse  \&
  Meylan}{Bonvin et~al.}{2016}]{bonvin2016cosmograil}
Bonvin V.,  Tewes M.,  Courbin F.,  Kuntzer T.,  Sluse D.,   Meylan G.,  2016,
  Astronomy \& Astrophysics, 585, A88

\bibitem[\protect\citeauthoryear{Bonvin et~al.,}{Bonvin
  et~al.}{2019}]{bonvin2019cosmograil}
Bonvin V.,  et~al., 2019, Astronomy \& Astrophysics, 629, A97

\bibitem[\protect\citeauthoryear{Braibant, Hutsemekers, Sluse  \&
  Goosmann}{Braibant et~al.}{2017}]{braibant2017constraining}
Braibant L.,  Hutsemekers D.,  Sluse D.,   Goosmann R.,  2017, Astronomy \&
  Astrophysics, 607, A32

\bibitem[\protect\citeauthoryear{Carbone, Castelli  \& Stanley}{Carbone
  et~al.}{2004}]{carbone2004analysis}
Carbone A.,  Castelli G.,   Stanley H.,  2004, Physical Review E, 69, 026105

\bibitem[\protect\citeauthoryear{Cornachione \& Morgan}{Cornachione \&
  Morgan}{2020}]{cornachione2020quasar}
Cornachione M.~A.,  Morgan C.~W.,  2020, The Astrophysical Journal, 895, 93

\bibitem[\protect\citeauthoryear{Ellingson, Yee  \& Green}{Ellingson
  et~al.}{1991}]{ellingson1991quasars}
Ellingson E.,  Yee H.,   Green R.,  1991, The Astrophysical Journal, 371, 49

\bibitem[\protect\citeauthoryear{Eulaers et~al.,}{Eulaers
  et~al.}{2013}]{eulaers2013cosmograil}
Eulaers E.,  et~al., 2013, Astronomy \& Astrophysics, 553, A121

\bibitem[\protect\citeauthoryear{Fisher, Bahcall, Kirhakos  \&
  Schneider}{Fisher et~al.}{1996}]{fisher1996galaxy}
Fisher K.~B.,  Bahcall J.~N.,  Kirhakos S.,   Schneider D.~P.,  1996, arXiv
  preprint astro-ph/9602078

\bibitem[\protect\citeauthoryear{Fohlmeister et~al.,}{Fohlmeister
  et~al.}{2007}]{fohlmeister2007time}
Fohlmeister J.,  et~al., 2007, The Astrophysical Journal, 662, 62

\bibitem[\protect\citeauthoryear{Gu \& Zhou}{Gu \&
  Zhou}{2010}]{gu2010detrending}
Gu G.-F.,  Zhou W.-X.,  2010, Physical Review E, 82, 011136

\bibitem[\protect\citeauthoryear{Guerras, Mediavilla, Jimenez-Vicente,
  Kochanek, Mu{\~n}oz, Falco  \& Motta}{Guerras
  et~al.}{2013}]{guerras2013microlensing}
Guerras E.,  Mediavilla E.,  Jimenez-Vicente J.,  Kochanek C.,  Mu{\~n}oz J.,
  Falco E.,   Motta V.,  2013, The Astrophysical Journal, 764, 160

\bibitem[\protect\citeauthoryear{Halsey, Jensen, Kadanoff, Procaccia  \&
  Shraiman}{Halsey et~al.}{1986}]{halsey1986fractal}
Halsey T.~C.,  Jensen M.~H.,  Kadanoff L.~P.,  Procaccia I.,   Shraiman B.~I.,
  1986, Physical review A, 33, 1141

\bibitem[\protect\citeauthoryear{Hawkins}{Hawkins}{2010}]{hawkins2010time}
Hawkins M.,  2010, Monthly Notices of the Royal Astronomical Society, 405, 1940

\bibitem[\protect\citeauthoryear{Hawkins}{Hawkins}{2020a}]{hawkins2020signature}
Hawkins M.,  2020a, Astronomy \& Astrophysics, 633, A107

\bibitem[\protect\citeauthoryear{Hawkins}{Hawkins}{2020b}]{hawkins2020sdss}
Hawkins M.,  2020b, Astronomy \& Astrophysics, 643, A10

\bibitem[\protect\citeauthoryear{Hawkins}{Hawkins}{2022}]{hawkins2022new}
Hawkins M.,  2022, Monthly Notices of the Royal Astronomical Society, 512, 5706

\bibitem[\protect\citeauthoryear{Hurst}{Hurst}{1951}]{hurst1951long}
Hurst H.~E.,  1951, Transactions of the American society of civil engineers,
  116, 770

\bibitem[\protect\citeauthoryear{Hutsem{\'e}kers, Braibant, Sluse, Anguita  \&
  Goosmann}{Hutsem{\'e}kers et~al.}{2017}]{hutsemekers2017new}
Hutsem{\'e}kers D.,  Braibant L.,  Sluse D.,  Anguita T.,   Goosmann R.,  2017,
  Frontiers in Astronomy and Space Sciences, 4, 18

\bibitem[\protect\citeauthoryear{Jim{\'e}nez-Vicente, Mediavilla, Kochanek,
  Munoz, Motta, Falco  \& Mosquera}{Jim{\'e}nez-Vicente
  et~al.}{2014}]{jimenez2014average}
Jim{\'e}nez-Vicente J.,  Mediavilla E.,  Kochanek C.,  Munoz J.,  Motta V.,
  Falco E.,   Mosquera A.,  2014, The Astrophysical Journal, 783, 47

\bibitem[\protect\citeauthoryear{Jovanovi{\'c}, Zakharov, Popovi{\'c}  \&
  Petrovi{\'c}}{Jovanovi{\'c} et~al.}{2008}]{jovanovic2008microlensing}
Jovanovi{\'c} P.,  Zakharov A.,  Popovi{\'c} L.,   Petrovi{\'c} T.,  2008,
  Monthly Notices of the Royal Astronomical Society, 386, 397

\bibitem[\protect\citeauthoryear{Kantelhardt, Zschiegner, Koscielny-Bunde,
  Havlin, Bunde  \& Stanley}{Kantelhardt
  et~al.}{2002}]{kantelhardt2002multifractal}
Kantelhardt J.~W.,  Zschiegner S.~A.,  Koscielny-Bunde E.,  Havlin S.,  Bunde
  A.,   Stanley H.~E.,  2002, Physica A: Statistical Mechanics and its
  Applications, 316, 87

\bibitem[\protect\citeauthoryear{{Kostrzewa-Rutkowska}
  et~al.,}{{Kostrzewa-Rutkowska} et~al.}{2018}]{kostrzewa2018gravitationally}
{Kostrzewa-Rutkowska} Z.,  et~al., 2018, \mn@doi [\mnras]
  {10.1093/mnras/sty259}, \href
  {https://ui.adsabs.harvard.edu/abs/2018MNRAS.476..663K} {476, 663}

\bibitem[\protect\citeauthoryear{Kumar et~al.,}{Kumar
  et~al.}{2013}]{kumar2013cosmograil}
Kumar S.~R.,  et~al., 2013, Astronomy \& Astrophysics, 557, A44

\bibitem[\protect\citeauthoryear{McLure \& Dunlop}{McLure \&
  Dunlop}{2001}]{mclure2001cluster}
McLure R.,  Dunlop J.,  2001, Monthly Notices of the Royal Astronomical
  Society, 321, 515

\bibitem[\protect\citeauthoryear{Millon et~al.,}{Millon
  et~al.}{2020a}]{millon2020cosmograil}
Millon M.,  et~al., 2020a, Astronomy \& Astrophysics, 640, A105

\bibitem[\protect\citeauthoryear{{Millon} et~al.,}{{Millon}
  et~al.}{2020b}]{millon2020cosmgrail}
{Millon} M.,  et~al., 2020b, \mn@doi [\aap] {10.1051/0004-6361/202037740},
  \href {https://ui.adsabs.harvard.edu/abs/2020A&A...640A.105M} {640, A105}

\bibitem[\protect\citeauthoryear{{Morgan}, {Kochanek}, {Morgan}  \&
  {Falco}}{{Morgan} et~al.}{2010}]{morgan2010quasar}
{Morgan} C.~W.,  {Kochanek} C.~S.,  {Morgan} N.~D.,   {Falco} E.~E.,  2010,
  \mn@doi [\apj] {10.1088/0004-637X/712/2/1129}, \href
  {https://ui.adsabs.harvard.edu/abs/2010ApJ...712.1129M} {712, 1129}

\bibitem[\protect\citeauthoryear{Mosquera \& Kochanek}{Mosquera \&
  Kochanek}{2011}]{mosquera2011microlensing}
Mosquera A.~M.,  Kochanek C.~S.,  2011, The Astrophysical Journal, 738, 96

\bibitem[\protect\citeauthoryear{Muzy, Bacry  \& Arneodo}{Muzy
  et~al.}{1991}]{muzy1991wavelets}
Muzy J.-F.,  Bacry E.,   Arneodo A.,  1991, Physical review letters, 67, 3515

\bibitem[\protect\citeauthoryear{Muzy, Bacry  \& Arneodo}{Muzy
  et~al.}{1994}]{muzy1994multifractal}
Muzy J.-F.,  Bacry E.,   Arneodo A.,  1994, International Journal of
  Bifurcation and Chaos, 4, 245

\bibitem[\protect\citeauthoryear{{Ochsenbein}, {Bauer}  \&
  {Marcout}}{{Ochsenbein} et~al.}{2000}]{vizier2000}
{Ochsenbein} F.,  {Bauer} P.,   {Marcout} J.,  2000, \mn@doi [\aaps]
  {10.1051/aas:2000169}, \href
  {https://ui.adsabs.harvard.edu/abs/2000A&AS..143...23O} {143, 23}

\bibitem[\protect\citeauthoryear{Popovi{\'c}, Mediavilla  \& Munoz}{Popovi{\'c}
  et~al.}{2001}]{popovic2001influence}
Popovi{\'c} L.,  Mediavilla E.,   Munoz J.,  2001, Astronomy \& Astrophysics,
  378, 295

\bibitem[\protect\citeauthoryear{Popovi{\'c}, Afanasiev, Shablovinskaya,
  Ardilanov  \& Savi{\'c}}{Popovi{\'c} et~al.}{2021}]{popovic2021spectroscopy}
Popovi{\'c} L.,  Afanasiev V.,  Shablovinskaya E.,  Ardilanov V.,   Savi{\'c}
  D.,  2021, Astronomy \& Astrophysics, 647, A98

\bibitem[\protect\citeauthoryear{Puckovs \& Matvejevs}{Puckovs \&
  Matvejevs}{2012}]{puckovs2012wavelet}
Puckovs A.,  Matvejevs A.,  2012, Information Technology \& Management Science
  (Sciendo)

\bibitem[\protect\citeauthoryear{Richardson, St{\"u}cker, Angulo  \&
  Hahn}{Richardson et~al.}{2022}]{richardson2022non}
Richardson T.,  St{\"u}cker J.,  Angulo R.,   Hahn O.,  2022, Monthly Notices
  of the Royal Astronomical Society, 511, 6019

\bibitem[\protect\citeauthoryear{{Shakura} \& {Sunyaev}}{{Shakura} \&
  {Sunyaev}}{1973}]{shakura1973black}
{Shakura} N.~I.,  {Sunyaev} R.~A.,  1973, \aap, \href
  {https://ui.adsabs.harvard.edu/abs/1973A&A....24..337S} {24, 337}

\bibitem[\protect\citeauthoryear{Shao, Gu, Jiang, Zhou  \& Sornette}{Shao
  et~al.}{2012}]{shao2012comparing}
Shao Y.-H.,  Gu G.-F.,  Jiang Z.-Q.,  Zhou W.-X.,   Sornette D.,  2012,
  Scientific reports, 2, 835

\bibitem[\protect\citeauthoryear{Shimizu, Thurner  \& Ehrenberger}{Shimizu
  et~al.}{2002}]{shimizu2002multifractal}
Shimizu Y.,  Thurner S.,   Ehrenberger K.,  2002, Fractals, 10, 103

\bibitem[\protect\citeauthoryear{Sluse, Hutsem{\'e}kers, Courbin, Meylan  \&
  Wambsganss}{Sluse et~al.}{2012}]{sluse2012microlensing}
Sluse D.,  Hutsem{\'e}kers D.,  Courbin F.,  Meylan G.,   Wambsganss J.,  2012,
  Astronomy \& Astrophysics, 544, A62

\bibitem[\protect\citeauthoryear{{Stone} et~al.,}{{Stone}
  et~al.}{2022}]{stone2022optical}
{Stone} Z.,  et~al., 2022, \mn@doi [\mnras] {10.1093/mnras/stac1259}, \href
  {https://ui.adsabs.harvard.edu/abs/2022MNRAS.514..164S} {514, 164}

\bibitem[\protect\citeauthoryear{Sun \& Malkan}{Sun \&
  Malkan}{1989}]{sun1989fitting}
Sun W.-H.,  Malkan M.~A.,  1989, The Astrophysical Journal, 346, 68

\bibitem[\protect\citeauthoryear{Taqqu, Teverovsky  \& Willinger}{Taqqu
  et~al.}{1995}]{taqqu1995estimators}
Taqqu M.~S.,  Teverovsky V.,   Willinger W.,  1995, Fractals, 3, 785

\bibitem[\protect\citeauthoryear{Telesca, Balasco, Colangelo, Lapenna  \&
  Macchiato}{Telesca et~al.}{2004}]{telesca2004investigating}
Telesca L.,  Balasco M.,  Colangelo G.,  Lapenna V.,   Macchiato M.,  2004,
  Physics and Chemistry of the Earth, Parts A/B/C, 29, 295

\bibitem[\protect\citeauthoryear{{Tewes}, {Courbin}  \& {Meylan}}{{Tewes}
  et~al.}{2013a}]{tewes2013cosmgrail9}
{Tewes} M.,  {Courbin} F.,   {Meylan} G.,  2013a, \mn@doi [\aap]
  {10.1051/0004-6361/201220123}, \href
  {https://ui.adsabs.harvard.edu/abs/2013A&A...553A.120T} {553, A120}

\bibitem[\protect\citeauthoryear{Tewes et~al.,}{Tewes
  et~al.}{2013b}]{tewes2013cosmograil}
Tewes M.,  et~al., 2013b, Astronomy \& Astrophysics, 556, A22

\bibitem[\protect\citeauthoryear{Wambsganss}{Wambsganss}{1998}]{wambsganss1998gravitational}
Wambsganss J.,  1998, Living Reviews in Relativity, 1, 1

\bibitem[\protect\citeauthoryear{Webster, Drinkwater  \& Thomas}{Webster
  et~al.}{1992}]{webster1992quasar}
Webster R.,  Drinkwater M.,   Thomas P.,  1992, in , Gravitational Lenses.
Springer, pp 230--236

\makeatother
\end{thebibliography}


\bsp	
\label{lastpage}
\end{document}